\documentclass[fleqn,usenatbib]{mnras}

\usepackage{newtxtext,newtxmath}
\usepackage[T1]{fontenc}

\DeclareRobustCommand{\VAN}[3]{#2}
\let\VANthebibliography\thebibliography
\def\thebibliography{\DeclareRobustCommand{\VAN}[3]{##3}\VANthebibliography}

\usepackage{graphicx}	% Including figure files
\usepackage{amsmath}	% Advanced maths commands
\usepackage{footnote}
\usepackage{threeparttable} % to use table notes
\usepackage{color}
\usepackage{caption}
\usepackage{CJK}
\usepackage[normalem]{ulem}

\newcommand{\RN}[1]{%
  \textup{\uppercase\expandafter{\romannumeral#1}}%
}

\title[UHECR-induced nuclear and EM cascades in radio galaxies]{ 
Nuclear and electromagnetic cascades induced by ultrahigh-energy cosmic rays in radio galaxies:
implications for Centaurus A}
\author[Zhang \& Murase]{
B. Theodore Zhang (张兵),$^{1}$\thanks{E-mail: bing.zhang@yukawa.kyoto-u.ac.jp}
Kohta Murase$^{2, 3, 4, 5, 1}$\thanks{E-mail: murase@psu.edu}
\\
$^1$Center for Gravitational Physics and Quantum Information, Yukawa Institute for Theoretical Physics, Kyoto University, Kyoto 606-8502, Japan\\
$^2$Department of Physics, The Pennsylvania State University, University Park, PA 16802, USA \\
$^3$Department of Astronomy \& Astrophysics, The Pennsylvania State University, University Park, PA 16802, USA \\
$^4$Center for Multimessenger Astrophysics, Institute for Gravitation and the Cosmos, The Pennsylvania State University, University Park, PA 16802, USA \\
$^5$School of Natural Sciences, Institute for Advanced Study, Princeton, NJ 08540, USA \\
}

\date{Accepted XXX. Received YYY; in original form ZZZ}

\pubyear{2022}

\begin{document}
\begin{CJK*}{UTF8}{gbsn}

\label{firstpage}
\pagerange{\pageref{firstpage}--\pageref{lastpage}}
\maketitle

\begin{abstract}
Very-high-energy (VHE) $\gamma$-rays ($\gtrsim 0.1\rm~TeV$) and neutrinos are crucial for identifying accelerators of ultrahigh-energy cosmic rays (UHECRs), but this is challenging especially for UHECR nuclei. In this work, we develop a numerical code to solve the transport equation for UHECRs and their secondaries, where both nuclear and electromagnetic cascades are taken into account self-consistently, considering steady UHECR accelerators such as radio galaxies. In particular, we focus on Centaurus A, which has been proposed as one of the most promising UHECR sources in the local universe. Motivated by observations of extended VHE $\gamma$-ray emission from its kiloparsec-scale jet by the H.E.S.S. telescope, we study interactions between UHECRs accelerated in the large-scale jet and various target photon fields including blazar-like beamed core emission, and present a quantitative study on VHE $\gamma$-ray signatures of UHECR nuclei, including the photodisintegration and Bethe-Heitler pair-production processes. We show that VHE $\gamma$-rays from UHECR nuclei could be detected by the ground-based $\gamma$-ray telescopes given that the dominant composition of UHECRs consists of intermediate-mass (such as oxygen) nuclei.
\end{abstract}


%%%%%%%%%%%%%%%%%%%%%%%%%%%%%%%%%%%%%%%%%%%%%%%%%%

%%%%%%%%%%%%%%%%% BODY OF PAPER %%%%%%%%%%%%%%%%%%

\section{Introduction}\label{set:intro}
Very-high-energy (VHE) $\gamma$-rays ($\gtrsim 0.1\rm~TeV$) from extragalactic sources have been detected by ground-based Imaging Atmospheric Cherenkov Telescopes (IACTs) and surface detectors~\citep[e.g.,][]{Hinton:2009zz, Sitarek:2022cpk}. 
Based on the unification model of radio-loud AGNs, blazars including BL Lac objects and flat-spectrum radio quasars (FSRQs) have relativistic jets pointing along the line-of-sight, while radio galaxies including Fanaroff-Riley (FR) I and II galaxies are viewed with an inclination angle to the jet axis where the effect of Doppler boosting is modest~\citep[e.g.,][]{Urry:1995mg}.
In the local universe, several radio-bright galaxies have been detected in the VHE band, including Centaurus A (Cen A), M87, NGC 1275, IC 310, 3C 264, and PKS 0625-35~\citep{rieger_radio_2018, Rulten:2022ekq}.
Cen A, the closest radio galaxy located at $\sim  3.7\rm~Mpc$ from Earth, has been extensively studied due to its potential as a source of ultrahigh-energy cosmic rays (UHECRs, $\gtrsim 1\rm~EeV$) detected~\citep[e.g.,][for reviews]{kotera_astrophysics_2011, anchordoqui_ultra-high-energy_2019, alves_batista_open_2019, Rieger:2022qhs}.
Additionally, the Pierre Auger Collaboration observed a $\sim 4\sigma$ deviation from isotropy at the intermediate angular scales in the Centaurus region for UHECRs with energies beyond $\sim 40\rm~EeV$~\citep{biteau_ultra-high-energy_2021}.
Cen A was also the first extragalactic extended source detected in the GeV sky by the \textit{Fermi}-LAT~\citep{the_fermi-lat_collaboration_fermi_2010}.
\cite{collaboration_resolving_2020} has provided strong evidence for VHE $\gamma$-ray production from components beyond the inner core of the radio galaxy~\citep{aharonian_discovery_2009, HESS:2018cxr}. 
However, current instruments still face limitations in resolving the emission regions from other VHE radio galaxies~\citep{Rulten:2022ekq}.

The production of VHE $\gamma$-rays from radio galaxies, such as Cen A, is thought to be the result of inverse-Compton (IC) emission by non-thermal electrons accelerated in the energy dissipation region within the jet.
This mechanism is consistent with the available evidence~\citep[e.g.,][]{collaboration_resolving_2020}.
However, hadronic scenarios, in which VHE $\gamma$-rays are produced through the interaction of high-energy cosmic rays, are still a viable possibility~\citep[e.g.,][]{petropoulou_one-zone_2014,fraija_gamma-ray_2014, fraija_neutrino_2016}. 
The hadronic components of the VHE $\gamma$-rays include electromagnetic cascade emission from photomeson production, hadronuclear interaction, Bethe-Heitler electron-position production, and photodisintegration accompanied by de-excitation $\gamma$-rays.
%~\citep[e.g.,][]{heiter_production_2018}. 
In magnetized environments, synchrotron radiation from UHECR protons and nuclei could also contribute significantly to the observed VHE $\gamma$-rays~\citep[e.g.,][]{Mucke:2000rn,Murase:2008mr}. 
The detection of de-excitation VHE $\gamma$-rays from nearby UHECR sources may provide direct evidence of the acceleration of the heavier nuclei components of UHECRs~\citep{murase_very-high-energy_2010}.
Based on observations showing that the fraction of heavier nuclei increases beyond $\sim 4$ EeV~\citep[e.g.,][]{guido_combined_2021}, it is likely that the highest energy range of UHECRs accelerated in nearby sources is dominated by heavier nuclei.

The photodisintegration of UHECR nuclei is a crucial process in determining their fate, especially when target photon energy in the nuclear rest frame exceeds the nuclear binding energy of approximately 10 MeV. This leads to the ejection of one or several nucleons, as described by \cite[][]{stecker_photodisintegration_1969}. Even fragmentation may occur when the target photon energy is high, and the photomeson production may occur when it exceeds the pion production threshold of approximately 140 MeV. As a result of photodisintegration, the nuclear fragments are left in an excited state, which quickly de-excites through the emission of one or several photons with energies around MeV in the nuclear rest frame. In the observer frame, these de-excitation $\gamma$-rays are boosted to the VHE range for ultra-relativistic cosmic ray nuclei.
The process was proposed for not only Galactic point sources~\citep[][]{karakula_gamma_1994, anchordoqui_tev_neutrino_2007, anchordoqui_tev_2007}
but also extragalactic sources~\citep{murase_very-high-energy_2010}.
It has been shown that UHECR nuclei can survive in radio galaxies like Cen A~\citep{Murase:2011cy}, implying that photodisintegration should not be efficient in regions where UHECRs are produced. 
For a given target photon energy the efficiency of de-excitation is also lower than those of the photomeson and Bethe-Heitler pair production, but TeV gamma rays can still be dominated by de-excitation $\gamma$-rays ~\citep[see, e.g., Fig.~1 of][]{murase_very-high-energy_2010}. 
The potential production of de-excitation VHE $\gamma$-rays from the core of Cen A via photodisintegration of heavy nuclei has been discussed~\citep{murase_very-high-energy_2010,kundu_photo-disintegration_2014,Leonel_Morejon_2021_ICRC}.

In this study, we present the numerical framework implemented in the Astrophysical Multimessenger Emission Simulator ({\sc AMES}). The framework can simultaneously treat both the nuclear cascade and the electromagnetic cascade by solving the coupled transport equations. %using the first-order implicit scheme.
The fate of UHECR nuclei in various astrophysical sources has been widely explored in previous studies~\citep[e.g.,][]{rodrigues_neutrinos_2018, zhang_low-luminosity_2018, biehl_cosmic_2018, boncioli_common_2019, Zhang:2018agl}. 
Our goal is to examine the impact of electromagnetic cascades on the multi-wavelength spectral energy distribution (SED), considering the injection of UHECR nuclei.
We apply our code to model leptohadronic processes in the large-scale jet of Cen A, where we investigate the detectability of de-excitation $\gamma$-rays. In contrast to the previous work on blazars, we consider the acceleration zone located in the large-scale jet, motivated by recent studies from the H.E.S.S. Collaboration~\citep[][]{collaboration_resolving_2020}. 
Additionally, we take into account the beamed photons from the inner core as the dominant target photons~\citep[e.g.,][]{bednarek_gevtev_2019, sudoh_physical_2020}.

The structure of the paper is as follows:
In Sec.~\ref{sec:fate}, we present an in-depth explanation of the physical processes related to the modeling of nuclear and electromagnetic cascades.
In Sec.~\ref{sec:application}, we use our numerical framework to investigate the hadronic origin of the VHE $\gamma$-rays detected from Cen A and the feasibility of detecting de-excitation $\gamma$-rays with present and future ground-based $\gamma$-ray detectors.
Sec.~\ref{sec:dis} explores the implications of our results.
Finally, in Sec.~\ref{sec:sum}, we summarize the work.

\section{Physical processes related to nuclear and electromagnetic cascade}\label{sec:fate}
To begin, let us consider a basic physical model where all physical processes occur within a uniform spherical emission region with comoving radius $l_b$ and radius from the black hole $R$. This emission region is encompassed by tangled magnetic fields with magnetic field strength $B$, and Doppler factor $\delta_D$.
In the following, we adopt the notation $E = \delta_D \varepsilon$, where $E$ represents the particle energy measured in the observer frame and $\varepsilon$ represents the particle energy measured in the comoving frame.
Note the redshift evolution factor $(1 + z)$ is not included considering the source is nearby. 

We assume the injection of CR nuclei following a power-law distribution with an exponential cutoff,
\begin{equation}
q(\varepsilon_A, t) \equiv \frac{dN^A}{dt d\varepsilon_A} = \dot{N}_{0}^A \left(\frac{\varepsilon_A}{Z \varepsilon_A^0} \right)^{-s_{\rm acc}} {{\rm exp}\left(-\frac{\varepsilon_A}{Z \varepsilon_{p}^{\rm max}} \right)},
\end{equation}
where $A$ is the CR nuclear mass number, $Z$ is the charge number, $s_{\rm acc}$ is the acceleration spectral index, $\varepsilon_{p}^{\rm max}$ is the proton maximum acceleration energy measured in the comoving frame, $\dot{N}_{0}^A$ is the normalization constant with units [$\rm eV^{-1} \ s^{-1}$], which is determined by
\begin{equation}
L_{\rm CR}=\frac{3}{2}\Gamma^2\sum_A \int d \varepsilon_A \frac{dN^A}{dt d\varepsilon_A},
\end{equation}
where $L_{\rm CR}$ is the total CR injection luminosity measured in the source frame.

We model the leptohadronic processes by solving a series of coupled transport equations for various particles including photons, electrons, neutrinos, neutrons, protons, and nuclei,
\begin{align}
\label{eq:transport}
\frac{\partial n_{\varepsilon_a}^a}{\partial t} &= -n_{\varepsilon_a}^a \mathcal{A}_a(\varepsilon_a) \nonumber \\ &+ \int d\varepsilon_a^* n_{\varepsilon_a^*}^a \mathcal{B}_{a \to a}(\varepsilon_a, \varepsilon_a^*) \nonumber \\ &+ \sum_b \int d\varepsilon_b n_{\varepsilon_b}^b \mathcal{C}_{b \to a}(\varepsilon_a, \varepsilon_b) \nonumber \\ &+ \dot{n}_a^{\rm inj}(\varepsilon_a),
\end{align}
where $a$ is particle type, $n_{\varepsilon_a}^a$ is particle differential number density at energy $\varepsilon_a$, $\mathcal{A}_a(\varepsilon_a)$ is the total interaction rate at energy $\varepsilon_a$ including particle escape, $\mathcal{B}_{a \to a}(\varepsilon_a, \varepsilon_a^*)$ is the self-production rate of particles with energy $\varepsilon_a$ generated from the same type of particles with energy $\varepsilon_a^*$, and $\mathcal{C}_{b \to a(\varepsilon_a, \varepsilon_b)}$ is the generation rate of particles with energy $\varepsilon_a$ from other types of particles with energy $\varepsilon_b$, and $\dot{n}_a^{\rm inj}$ is the source injection rate at energy $\varepsilon_a$. 
In Appendix~\ref{app:solution}, we provide the details of the numerical scheme and the related physical processes described in the above sections for each particle species, respectively. We here provide a part of the Astrophysical Multimessenger Emission Simulator (AMES) code, which integrates those for hadronic and leptonic emissions from various astrophysical objects. Some of the earlier calculations without nuclear cascades are found in, e.g., \cite{M12,MB12,Murase:2014bfa,murase_new_2018,Murase:2022dog}. 

The photonuclear interaction rate for CR nuclei can be calculated by
\begin{align}\label{eq:interaction-rate}
    t_{A\gamma}^{-1} &= c \oint d\Omega \int_0^\infty d\varepsilon (1 - \beta_A \mu) \frac{dn}{d\varepsilon d\Omega} \sigma_{A\gamma} (\bar{\varepsilon}) \nonumber \\
    &= \frac{c}{2} \int_{-1}^{1} d\mu \int_0^\infty d\varepsilon (1 - \beta_A \mu) \frac{dn}{d\varepsilon} \sigma_{A\gamma} (\bar{\varepsilon}),
\end{align}
where $c$ is the speed of light, the solid angle averaged target photon distribution $dn/d\varepsilon \equiv  (1/4\pi) \oint d\Omega (dn/d\varepsilon d\Omega)$, $\beta_A$ is the particle velocity, $\mu = \rm cos \theta$ is the angle between nuclei and incident target photon, $\sigma_{A\gamma}$ is the cross section, and $\bar{\varepsilon} = \gamma_A (1 - \beta_A \mu) \varepsilon$ is the target photon energy measured in the nuclear rest frame. 
We can define optical depth as
\begin{equation}
    \tau_{A\gamma} \approx t_{\rm esc} / t_{A\gamma},
\end{equation}
where $t_{\rm esc}$ is the escape time and in the limit that it is dominated by advection we have 
\begin{equation}
    t_{\rm esc} \approx t_{\rm adv} \approx \frac{l_b}{V},
\end{equation}
where $t_{\rm adv}$ is the advection time scale.

Note that the energy loss rate, which is useful in analytical estimates, can be calculated similarly, by considering the inelasticity $\kappa_{A\gamma}$.

For the purpose of understanding physical processes analytically, let us consider target photon fields with a broken power law,
\begin{equation}\label{eq:target}
    \frac{dn}{d\varepsilon} = n_{\varepsilon_{b}}
    \begin{cases} (\varepsilon / \varepsilon_{b})^{-\alpha_l}, \varepsilon < \varepsilon_{b} \\ (\varepsilon / \varepsilon_{b})^{-\alpha_h}, \varepsilon > \varepsilon_{b}
    \end{cases},
\end{equation}
in the comoving frame of the blob, where $n_{\varepsilon_{b}}$ is the differential photon energy density at $\varepsilon_{b}$ with units [$\rm eV^{-1}~cm^{-3}$], $\varepsilon_{b}$ is the break energy, $\alpha_l$ and $\alpha_h$ are spectral indices. 

\begin{figure*}
    \includegraphics[width=0.7\linewidth]{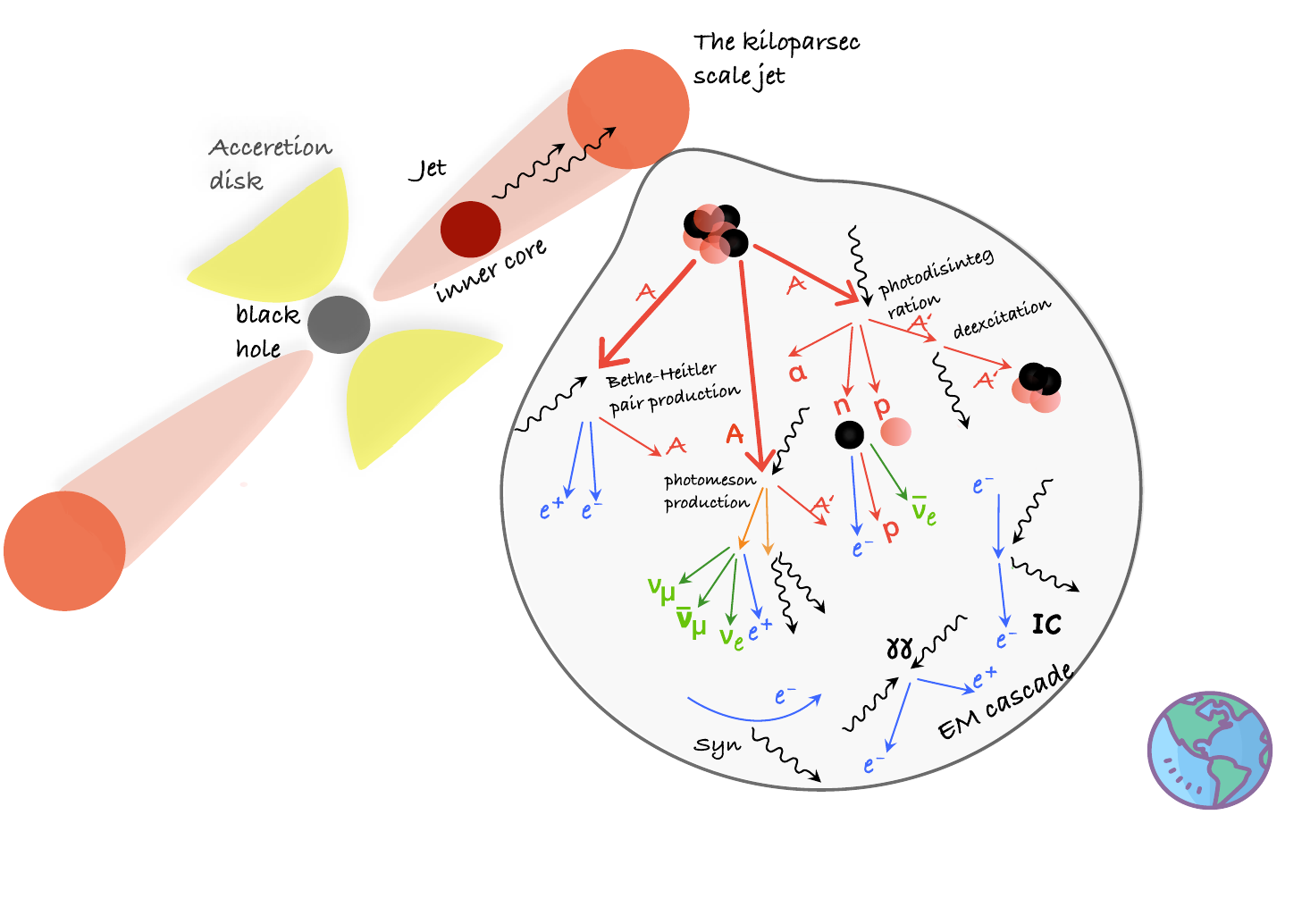}
    \caption{Schematic picture of neutrino and gamma-ray production processes caused by protons, nuclei and electrons, including nuclear and electromagnetic cascades induced by CR nuclei in radio galaxies.}
    \label{fig:cenA}
\end{figure*}

In Fig.~\ref{fig:cenA}, we show a schematic picture of interactions by CR nuclei and electrons, including nuclear and electromagnetic cascades induced by CR nuclei, in the large-scale jet of radio galaxies. The relevant physical processes include photodisintegration of nuclei, photomeson production process, Bethe-Heitler pair production process, and accompanied electromagnetic cascades.

\subsection{Photodisintegration process}
The photodisintegration process is inelastic interaction between CR nuclei and target photons, resulting in the emission of one or more nucleons or lighter nuclei from a parent nuclei, as
\begin{equation}
 A + \gamma \to A_d^\star + {\rm nucleons}, 
\end{equation}
where $A$ represents a primary nuclei, $\gamma$ is an incoming target photon and $A_d^\star$ is a daughter nuclei. 
This process occurs when the incoming target photons in the nuclear rest frame have energy greater than the nuclear binding energy, typically $\bar{\varepsilon}_\gamma \gtrsim 10\rm~MeV$~\citep[e.g.,][]{Rachen:1996zeh}.
The photodisintegration process is typically dominated by the giant dipole resonance (GDR), which may be approximated by the $\delta$-function,
$\sigma_{\rm phdis} \approx \sigma_{\rm GDR} \delta(\bar{\varepsilon}_\gamma - \bar{\varepsilon}_{\rm GDR})$, 
where $\bar{\varepsilon}_{\rm GDR}$ is the typical photon energy in the nuclear rest frame, and the width of the GDR process is $\Delta \bar{\varepsilon}_{\rm GDR}$.
%Insert Eq.~\ref{eq:cross} into Eq.~\ref{eq:rate}, 
The optical depth to the photodisintegration process is estimated by~\citep[e.g.,][]{murase_very-high-energy_2010},
\begin{align}
\label{eq:opticaldepth-inte-dis}
\tau_{A\gamma} &\approx t_{\rm esc} / t_{\rm phdis} \nonumber \\ 
&\approx\frac{2}{\alpha+1} (c/V) l_b \hat{\sigma}_{\rm dis} \varepsilon_{b} n_{\varepsilon_{b}}  \left(\frac{E_A}{E_{A, b}^{\rm dis} }\right)^{\alpha -1} \nonumber\\
&\sim 0.05 \frac{2}{\alpha+1} \left(\frac{A}{16}\right)^{1.21} \left(\frac{V}{0.5 c}\right)^{-1} \left(\frac{l_b}{1\rm~kpc}\right) \left(\frac{E_A}{E_{A, b}^{\rm dis}}\right)^{\alpha -1},
\end{align}
 where $t_{\rm phdis}$ is the photodisintegration interaction time scale, $\varepsilon_{b} n_{\varepsilon_{b}} = 10^3\rm~cm^{-3}$, $\varepsilon_{b} = 10\rm~eV$, $\hat{\sigma}_{\rm dis} = \kappa_{\rm GDR} \sigma_{\rm GDR} \Delta \bar{\varepsilon}_{\rm GDR} / \bar{\varepsilon}_{\rm GDR}$ is the effective photo disintegration cross section, $\kappa_{\rm GDR}$ is the inelasticity, $\sigma_{\rm GDR} \approx 1.45\times 10^{-27} A\rm~cm^2$, $\bar{\varepsilon}_{\rm GDR} \approx 42.65 A^{-0.21}\rm~MeV$ ($A > 4$), $\Delta \bar{\varepsilon}_{\rm GDR} \approx 8\rm~MeV$ and $E_{A,b}^{\rm dis} \approx 0.5  \delta_D m_A c^2 \bar{\varepsilon}_{\rm GDR} / \varepsilon_{b} \simeq 7.2 \times 10^{16}~\delta_D (A / 16)^{0.79} (\varepsilon_b / 10\rm~eV)^{-1}\rm~eV$ is the typical energy of the nuclei that interacts with target photons with energy $\varepsilon_b$.
For the photodisintegration process with the ejection of one nucleon, we have $\kappa_{\rm GDR} = 1 / A$. However, in general, the mean inelasticity depends on the contributions from all the dominant channels, see the details in Appendix A of ~\cite{zhang_high-energy_2017}.

The corresponding effective optical depth is
\begin{align}\label{eq:opticaldepth-dis}
    f_{\rm dis} &\approx t_{\rm esc}/t_{\rm dis}
    \sim \tau_{A\gamma}/A \nonumber\\
    &\sim 3 \times 10^{-3} \frac{2}{\alpha+1} \left(\frac{A}{16}\right)^{0.21} \left(\frac{V}{0.5 c}\right)^{-1} \left(\frac{l_b}{1\rm~kpc}\right) \left(\frac{E_A}{E_{A, b}^{\rm dis}}\right)^{\alpha -1},
\end{align}
where $t_{\rm dis}$ is the energy loss time scale.

The remaining daughter nuclei are in an excited state, which undergoes a subsequent de-excitation process by emitting one or multiple photons,
\begin{equation}
    A_d^\star \to A_d + {\rm photons},
\end{equation}
where $\bar{\varepsilon}_{\gamma, \rm deex}$ is the typical photon energy in the nuclear rest frame.
The value of $\bar{\varepsilon}_{\gamma, \rm deex}$ extends from hundreds keV to a few MeV, which depends on the specific de-excitation channel~\citep{anchordoqui_tev_2007,murase_very-high-energy_2010, kundu_photo-disintegration_2014}.
The observed de-excited $\gamma$-ray have boosted energy $E_{\gamma, \rm deex} = \delta_D \gamma_A \bar{E}_{\gamma, \rm deex} \sim 100 \delta_D (\gamma_A / 10^8) (\bar{E}_{\gamma, \rm deex} / 1\rm~MeV)\rm~TeV$, where $\gamma_A$ is the Lorentz factor of the daughter nuclei $A_d^\star$.
The detailed inclusive cross section for the production of de-excitation photons can be calculated with numerical code {\sc Talys}~\citep{goriely_improved_2008}.
However, the direct output of the total cross-section of the photodisintegration process for light and intermediate-mass nuclei is inconsistent with experimental data~\citep{batista_crpropa_2016, tamii_pandora_2022}.
For this purpose, we directly adopt the same data files used in {\sc CRPropa 3} which had been prepared with {\sc Talys 1.8} using adjusted GDR parameters in order to better match the experimental data~\citep{batista_crpropa_2016}.
However, considering uncertainties and the discrepancies in the photodisintegration process for the inclusive cross section for photon production of {\sc Talys} with other numerical code, e.g., {\sc Fluka}~\citep{bohlen_fluka_2014}, we also adopt the method used in previous works (e.g., \cite{anchordoqui_tev_2007, murase_very-high-energy_2010, kundu_photo-disintegration_2014}).
For numerical calculations, the injection rate of de-excitation $\gamma$-rays from the photodisintegration process is calculated using Eq.~16 of \citet{anchordoqui_tev_2007}.
We assume the average energy of emitted photons is $\bar{\varepsilon}_{\gamma, \rm deex} = 2\rm~MeV$ when measured in the nuclear rest frame and the multiplicity is $\bar{n}_\gamma = 3$~\citep{Morejon:2021nvz}.
The effective optical depth is estimated to be~\citep{murase_very-high-energy_2010}
\begin{align}\label{eq:opticaldepth-deex}
f_{\rm deex} &\approx t_{\rm esc} / t_{\rm deex} \approx (\kappa_{\rm deex} / \kappa_{\rm GDR}) f_{\rm dis} \nonumber \\ &\sim 2\times 10^{-5} \frac{2}{\alpha+1} \left(\frac{A}{16}\right)^{0.21} \left(\frac{V}{0.5 c}\right)^{-1} \left(\frac{l_b}{1\rm~kpc}\right) \left(\frac{E_A}{E_{A, b}^{\rm dis}}\right)^{\alpha -1},
\end{align}
where $\kappa_{\rm deex} \approx \bar{n}_\gamma \bar{\varepsilon}_{\gamma, \rm deex} / m_A c^2$ is the energy carried by the de-excitation $\gamma$-rays.
Furthermore, we ignore the de-excitation photons generated from the excited fragments via photomeson production.

\subsection{Photomeson production process}
The photomeson production process occurs when the energy of target photons in the nuclear rest frame exceeds the pion production threshold, $\varepsilon_{\rm th} \approx m_{\pi} c^2 (1 + m_{\pi} / 2 m_N) \sim 140\rm~MeV$.
The effective optical depth of the photomeson production process can be estimated using the $\Delta$-resonance as
\begin{align}\label{eq:opticaldepth-mes}
    f_{\rm mes} &\approx t_{\rm esc}/t_{\rm mes} \sim \frac{2}{\alpha+1} (c/V) l_b \hat{\sigma}_{\rm mes} \varepsilon_{b} n_{\varepsilon_{b}}   \left(\frac{E_A}{E_{A, b}^{\rm mes}}\right)^{\alpha -1} \nonumber \\  &\sim 0.3 \times 10^{-3} \frac{2}{\alpha+1} \left(\frac{V}{0.5 c}\right)^{-1} \left(\frac{l_b}{1\rm~kpc}\right) \left(\frac{E_A}{E_{A, b}^{\rm mes}}\right)^{\alpha -1},
\end{align}
where $t_{\rm mes}$ is the photomeson production energy loss time scale, $\hat{\sigma}_{\rm mes} \sim A \hat{\sigma}_{p\gamma} \approx A \kappa_{\rm mes} \sigma_{\Delta} \Delta \bar{\varepsilon}_{\Delta} / \bar{\varepsilon}_{\Delta}$, $\kappa_{\rm mes}\sim \kappa_{p\gamma}/A$ is the nuclear inelasticity, $\kappa_{p\gamma} \sim 0.2$ is the proton inelasticity, $\sigma_{\Delta} \approx 4.4\times 10^{-28}\rm~cm^2$, $\bar{\varepsilon}_{\Delta} \simeq 0.34\rm~GeV$, $\Delta \bar{\varepsilon}_{\Delta} \simeq 0.2\rm~GeV$ and $E_{A,b}^{\rm mes} \approx 0.5 \delta_D m_A c^2 \bar{\varepsilon}_{\Delta} / \varepsilon_{b} \simeq 2.5\times 10^{17} \delta_D (A / 16) (\varepsilon_b / 10\rm~eV)^{-1}\rm~eV$~\citep[e.g.,][]{murase_very-high-energy_2010}.
The photomeson production interaction time scale $t_{\rm phmes}$ can be estimated without considering the inelasticity.
In our numerical approach, we employ the Monte Carlo event generator {\sc SOPHIA} to determine the differential cross-section of all stable secondary particles~\citep{Mucke:1999yb}.
We adopt the superposition model, where the photomeson cross section is $\sigma_{\rm mes} \sim A \sigma_{p\gamma}$, where $\sigma_{p\gamma}$ is the photomeson cross-section of protons~\citep{Zhang:2018agl}. 
The superposition model is employed in {\sc CRPropa 3} with a slightly different scaling law as described in Eq. 3 in~\cite{kampert_crpropa_2013}, which assumes the emission of one proton or neutron in each interaction process.
In accordance with~\cite{batista_crpropa_2016}, we assume that the cross-section for nuclei with the mass number $A > 8$ is $0.85 (Z + N) \sigma_{p\gamma}$ times higher, where $Z$ and $N$ are the numbers of protons and neutrons, respectively. 
On the other hand, for nuclei with the mass number $1 < A \leqslant 8$, the cross section is estimated to be $0.85 (Z^{2/3} + N^{2/3}) \sigma_{p\gamma}$.
In~\cite{morejon_improved_2019}, a comprehensive examination of the photomeson production process that incorporates the impact of the nuclear medium and the fragmentation of the primary nucleus is presented.
In the context of radio galaxies under consideration in this work, the target photons typically have lower energies and the photodisintegration process of primary nuclei is dominant, thus we opt for the simplified superposition model for ease of calculation.
However, the photomeson production process and nuclear fragmentation become significant when the target photons are produced by the prompt emission of relativistic jets, such as in $\gamma$-ray bursts and tidal disruption events~\citep{Murase:2008mr,murase_very-high-energy_2010,morejon_improved_2019}.

\subsection{Bethe-Heitler pair production process}
The Behte-Heitler pair production process will occur once the target photons have energy beyond $\varepsilon_{\rm th} \approx 2 m_e c^2 \sim 1\rm~MeV$ in the nuclear rest frame~\citep[e.g.,][]{Blumenthal:1970nn}.
The effective optical depth to the Bethe-Heitler pair production process is estimated to be
\begin{align}\label{eq:opticaldepth-BH}
    f_{\rm BH} &\approx t_{\rm esc}/t_{\rm BH} \sim \frac{2}{\alpha+1} (c/V) l_b \hat{\sigma}_{\rm BH}\varepsilon_{b} n_{\varepsilon_{b}}  \left(\frac{E_A}{E_{A, b}^{\rm BH}}\right)^{\alpha -1} \nonumber \\  &\sim 2 \times 10^{-5} \frac{2}{\alpha+1}
 \left(\frac{Z}{8}\right)^{2} \left(\frac{A}{16}\right)^{-1} \left(\frac{V}{0.5 c}\right)^{-1} \left(\frac{l_b}{1\rm~kpc}\right) \left(\frac{E_A}{E_{A, b}^{\rm BH}}\right)^{\alpha -1},
\end{align}
where $\hat{\sigma}_{\rm BH} \sim 8 \times 10^{-31} (Z^2/A) \rm~cm^{2}$ is the pair production cross section of nuclei, $E_{A,b}^{\rm BH} \approx 0.5 \delta_D \bar{\varepsilon}_{\rm BH} m_A c^2 / \varepsilon_{b} \simeq 7.5\times10^{15}\delta_D (A / 16) (\varepsilon_b / 10\rm~eV)^{-1}\rm~eV$ and $\bar{\varepsilon}_{\rm BH} \approx 10\rm~MeV$.
To calculate the energy loss rate of the photopair production process for relativistic nuclei, we employ Equation 3.11 in~\cite{chodorowski_reaction_1992}. 
The energy spectrum of secondary electron-positron pairs can be determined by utilizing Eq. 62 in~\cite{kelner_energy_2008}. 
This equation calculates the double-differential cross-section of emitted electrons (and positrons) as a function of energy and emission angle in the nuclear rest frame, which is obtained from Eq. 10 in~\cite{blumenthal_energy_1970}. 
We note that a factor of 2 should be multiplied when using Eq. 62 in~\cite{kelner_energy_2008} to accurately account for both electrons and positrons.

Note that the relative contribution of de-excitation VHE $\gamma$-rays and the Bethe-Heitler pair production process to the observed flux at the TeV energy range depends on several factors, such as the source magnetic field strength, target photon fields, and the composition of UHECR nuclei, as noted in studies by~\citet{murase_very-high-energy_2010, aharonian_limitations_2010}.
The energy loss rate due to de-excitation $\gamma$-rays can be lower than that of the Bethe-Heitler pair production process, but the relative contribution depends on details of electromagnetic cascades from the Bethe-Heitler electron-positron pairs~\citep[e.g.,][]{murase_very-high-energy_2010}.

\subsection{Electromagnetic cascade}
High-energy electrons and positrons will lose energy through processes such as synchrotron emission and inverse-Compton (IC) scattering, which are influenced by the strength of the magnetic fields and the density of the ambient target photon fields, respectively. 
The synchrotron energy loss time scale for high-energy particles is
\begin{equation}
    t_{\rm syn}^{-1} \approx \frac{4 \sigma_T Z^4 m_e^2}{3 m^4 c^3} \frac{E_A}{\delta_D} U_B,
\end{equation}
where $Z$ is the particle charge, $m$ is the particle mass, $\sigma_T$ is the Thomson cross section, and $U_B = B^2/8\pi$ is the magnetic energy density~\citep[e.g.,][]{rybicki_radiative_1986}.
The IC energy loss time scale for high-energy electrons is given by
\begin{equation}
    t_{\rm IC}^{-1} \approx \frac{4 \sigma_T c}{3 m_e^2 c^4} \frac{E_e}{\delta_D} U_{\rm ph} F_{\rm KN},
\end{equation}
where $U_{\rm ph}$ is the energy density of target photons and $F_{\rm KN}$ is a factor to take into account Klein-Nishina effect~\citep{Jones:1968zza}.
In our numerical calculations, we use the known results of the total and differential cross sections~\citep[e.g.,][]{rybicki_radiative_1986}. 
In this work, we also adopt the continuous energy loss approximation, which is further detailed in Appendix~\ref{app:solution}.

High-energy $\gamma$-rays may interact with target photons, leading to the creation of electron-positron pairs.
The two-photon annihilation optical depth can be estimated by
\begin{align}\label{eq:opticaldepth-gg}
    \tau_{\gamma \gamma} &\approx 
    t_{\rm lc}/t_{\rm \gamma \gamma} \sim \eta_{\gamma \gamma} \sigma_T l_b \varepsilon_{b} n_{\varepsilon_{b}} \left(\frac{E_\gamma}{E_{\gamma, b}}\right)^{\alpha - 1} \nonumber \\ &\sim 0.2 \left(\frac{l_b}{1\rm~kpc}\right) \left(\frac{E_\gamma}{E_{\gamma, b}}\right)^{\alpha - 1},
\end{align}
where $t_{\gamma\gamma}$ is the two-photon annihilation time scale, $\eta_{\gamma \gamma} \sim 0.1$ is a numerical factor that depends on the target photon spectral index~\citep[e.g.,][]{svensson_non-thermal_1987}, $E_{\gamma, b} \approx \delta_D (m_e c^2)^2 / \varepsilon_{b} \simeq 2.6\times10^{10} \delta_D (\varepsilon_b / 10)^{-1} \rm~eV$. 
Note that the validity of Eq.~\ref{eq:opticaldepth-gg} is based on the assumption that the target photon spectrum is soft, with $\alpha \gtrsim 1$~\citep[e.g.,][]{dermer_high_2009}. 
In this work for simplicity we also assume that electrons and positrons share the photon energy \citep[but the detailed distribution may be used as in][]{MB12}.
The injection of high-energy $\gamma$-rays and electrons will trigger an electromagnetic cascade process, which is essential for predicting the observed multi-wavelength energy spectrum.

In addition to the isotropic IC scattering case, we also consider the effect of anisotropic IC scattering process~\citep[e.g.,][]{brunetti_anisotropic_2000}. 
It had been proposed that the observed flux from the anisotropic IC scattering process is sensitive to the observation angle, especially for nearby radio galaxies~\citep[e.g.,][]{brunetti_anisotropic_2000, bednarek_morphology_2020}.

\section{Application to the nearest radio galaxy Centaurus A}\label{sec:application}
\subsection{Physical model}
In this section, we apply the above method to the nearest radio galaxy, Cen A. 
Motivated by the possible connection between Cen A and the observed UHECRs~\citep[e.g.,][]{biteau_ultra-high-energy_2021}, the hadronic origin of the VHE $\gamma$-rays from Cen A has been explored by various authors based on the inner core model~\citep[e.g.,][]{fraija_gamma-ray_2014, petropoulou_one-zone_2014, fraija_study_2018, joshi_very_2018, fraija_study_2018,banik_interpreting_2020}. 
However, due to the smaller distance of the inner core to the central black hole, these models contradict the observed morphology of VHE $\gamma$-rays, which is consistent with the origin from the kiloparsec-scale jet~\citep{collaboration_resolving_2020}. 

It is natural to expect that charged particles, including electrons and nuclei, could accelerate in the kiloparsec-scale jet via stochastic acceleration or shear acceleration.
The accelerated high-energy electrons could up-scatter the surrounding target photon fields to the VHE energy range, such as infrared photons from dust torus~\citep{collaboration_resolving_2020, Liu:2017gln, wang_particle_2021}, optical and ultra-violet emission from disk starlight~\citep{hardcastle_modelling_2011, tanada_inverse_2019} and broad-band non-thermal emission from the inner core or ``hidden'' core~\citep{bednarek_gevtev_2019, bednarek_morphology_2020}.

Shear acceleration could operate when charged particles are scattered off the magnetic field inhomogeneities from different layers of the shearing flow~\citep[e.g.,][]{Rieger:2019hrf}.
In the steady-state, the energy spectrum of accelerated particles typically may follow a power-law distribution with an exponential cutoff $dN^A/d\varepsilon \propto \varepsilon^{-s_{\rm acc}} {\rm exp}(\varepsilon/\varepsilon_{\rm max})$, where $s_{\rm acc}$ is the spectral index. 
The spectral index $s_{\rm acc}$ of accelerated particles without radiation energy losses could be much steeper for non-relativistic flow speeds because of the efficient diffusive escape process~\citep{Rieger:2022oce}. On the other hand, the spectral index is harder for trans-relativistic and relativistic flows and the shear acceleration has been proposed as an effective mechanism to accelerate charged nuclei to the UHE energy range in the large-scale jet~\citep[e.g.,][]{kimura_ultrahigh-energy_2018,Wang:2022cnl,Rieger:2022qhs, Seo:2022zmn}.
For large-scale relativistic jets, not only the shear reacceleration but also the one-shot reacceleration can be efficient, especially for inner jets and powerful FR-II jets~\citep{Mbarek:2021bay,Mbarek:2022nat}.
The maximum available energy of the accelerated CR nuclei can be estimated under the Hillas-type confinement condition~\citep{hillas_origin_1984},
\begin{align}\label{eq:Emax}
    E_{\rm max} &\approx \frac{1}{\eta} \Gamma_{\rm kpc} Z e B \beta l_b \nonumber \\ &\simeq 110 (Z/26) {\Gamma_{\rm kpc}} \left(\frac{\eta}{7}\right)^{-1} \left( \frac{B}{10^{-4}\rm~G} \right) \left( \frac{\beta}{0.3} \right) \left( \frac{l_b}{1 \rm~kpc} \right) \rm~EeV,
\end{align}
where $\eta$ represents a prefactor which is $\sim$a few in the Bohm limit~\citep[e.g.,][]{Drury:1983zz}, $B$ is the magnetic field strength, $\beta$ is the shock velocity.
From Eq.~\ref{eq:Emax}, we can see the required magnetic luminosity is~\citep{2009PhRvD..80l3018P, Murase:2011cy}
\begin{equation}\label{eq:LB}
    L_B \sim 4 \times 10^{44} {(Z/26)}^{-2} \left(\frac{\eta}{7}\right)^{2} \left(\frac{E_{\rm max}}{100 \rm~EeV}\right)^2 \left( \frac{\beta}{0.3} \right)^{-1} \rm~erg~s^{-1}.
\end{equation}
The Eq.~\ref{eq:LB} gives the condition for a source capable of acceleration particles to 100 EeV.
Note that the composition of the accelerated UHECR nuclei could be different from the typical composition of the interstellar medium~\citep[e.g.,][]{kimura_ultrahigh-energy_2018}. 
In this work, for the demonstration, we only consider two typical elements of heavier nuclei, oxygen, and iron nuclei, to study the detectability of de-excitation VHE $\gamma$-rays. 

For simplicity, we assume the emission region in the kiloparsec-scale jet is modeled as a spherical blob moving at sub-relativistic speed towards us with the same inclination angle $\theta_{\rm ob}$ as the inner core~\citep{collaboration_resolving_2020}.
The viewing angle of the jet is $\theta_{\rm ob} \sim 20^\circ - 50^\circ$~\citep{tingay_subparsec-scale_1998, hardcastle_radio_2003}.
The jet velocity can be derived from observed apparent velocity $\beta_{\rm app}$ combining with the viewing angle $\theta_{\rm ob}$, $\beta = \beta_{\rm app} / (\beta_{\rm app} {\rm cos} \theta_{\rm ob} + {\rm sin} \theta_{\rm ob})$.
Similar to~\cite{bednarek_gevtev_2019, bednarek_morphology_2020}, we consider the broad-band non-thermal emission originating from the inner core as the dominant target photon field in the kiloparsec-scale jet.
The comoving frame photon energy density of the inner core emission ``observed'' in the kpc-scale jet is
\begin{equation}
    {\varepsilon}^2\frac{dn}{d\varepsilon} = \frac{d_L^2}{R^2 c} \frac{\delta_{D, \rm core, kpc}^4}{\delta_{D, \rm core, ob}^4} \delta_{D, \rm deboost}^4 E F_{ E},
\end{equation}
where $d_L$ is the luminosity distance to the observer at Earth, $R = 1\rm~kpc$ is the distance of the kiloparsec-scale jet to the inner core, $\delta_{D, \rm core, ob} = 1 / (\Gamma_{\rm core} (1 - \beta_{\rm core} {{\rm cos} \theta_{\rm ob}}))$ is the Doppler factor of the inner core when observed at Earth, $\delta_{D, \rm core, kpc} = 1 / (\Gamma_{\rm core} (1 - \beta_{\rm core}))$ is the Doppler factor of the inner core when observed at the distance of kpc-scale jet in the black hole rest frame, $\delta_{D, \rm deboost} = 1 / (\Gamma_{\rm kpc} (1 + \beta_{\rm kpc}))$ is the de-boosting factor when converting the inner core emission from black hole rest frame to the comoving frame, and $E F_{ E}$ is the observed SED of the inner core.
In this work, we adopt $\Gamma_{\rm core} = 5$ and $\Gamma_{\rm kpc} = 1.05$. If we adopt the jet viewing angle as $\theta_{\rm ob} = 40^\circ$, the corresponding Doppler factor are $\delta_{D, \rm core, ob} \simeq 1$, $\delta_{D, \rm core, kpc} \simeq 8$ and $\delta_{D, \rm deboost} \simeq 0.7$.
We fit the observed SED of inner core emission with the following formula,
\begin{equation}\label{eq:two-bump}
    E F_{E} = A x^{-\alpha_1} \left(\frac{1+x^{1/\alpha_3}}{2}\right)^{(\alpha_1 - \alpha_2) \alpha_3},
\end{equation}
where $x = E / E_b$. The adopted values for the low-energy bump are $A = 1.5 \times 10^{-10}\rm~erg~cm^{-2}~s^{-1}$, $E_b = 1\rm~eV$, $\alpha_1 = -0.8$, $\alpha_2 = 3.2$, and $\alpha_3 = 1$. The values for the high-energy bump are $A = 4.5 \times 10^{-10}\rm~erg~cm^{-2}~s^{-1}$, $E_b = 10^6\rm~eV$, $\alpha_1 = -0.6$, $\alpha_2 = 1$, and $\alpha_3 = 3$.

\subsection{Time Scales}
\begin{figure}
	\includegraphics[width=\linewidth]{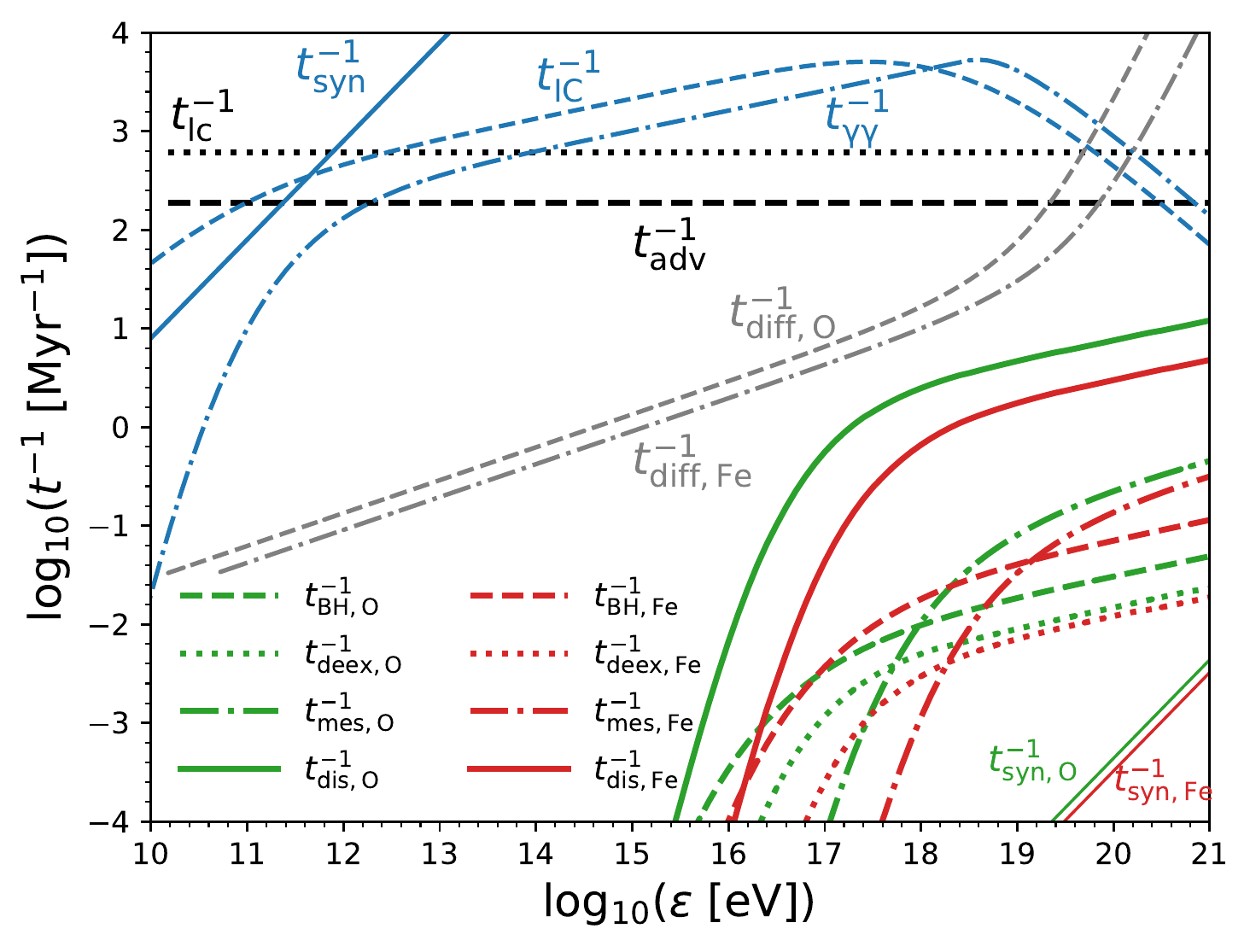}
    \caption{Different time scales of various energy loss processes in the kiloparsec-scale jet, with the light crossing time scale indicated as a dotted line, the advection time scale represented by the black dashed line and the diffusion escape time scales shown in grey lines. }
    \label{fig:result-timescale}
\end{figure}

In Fig.~\ref{fig:result-timescale}, we show various time scales in the kiloparsec-scale jet.
We model the kiloparsec-scale jet as a spherical blob located at a distance $r_{\rm kpc} = 1\rm~kpc$ from the inner core.
The blob has a comoving radius of $l_b = 0.5\rm~kpc$ and moves with a velocity of $V = \beta_{\rm kpc} c = 0.3$ towards the observer with a viewing angle of $\theta_{\rm ob} \sim 40\rm~deg$. The Doppler factor is $\delta_{D,\rm kpc}$.
To simplify calculations, we assume isotropic target photon fields are uniformly distributed throughout the kiloparsec-scale jet region.
However, in reality, the emission from the inner core is highly beamed and concentrated along the jet axis, varying depending on the activity of the central supermassive black holes.
The mean-field magnetic field strength is set to $B = 1\times 10^{-4}\rm~G$. 

The black-dashed line represents the advection time scale $t_{\rm adv} = l_b / V$ for charged particles.
We also consider the diffusive escape process for oxygen nuclei (grey dashed line) and iron nuclei (grey dot-dashed line). 
The diffusive escape time scales can be estimated by $t_{\rm diff} \approx l_b^{2} / 6D$ assuming spherical geometry, where $D$ is the diffusion coefficient.
We adopt the following form of the diffusion coefficient
\begin{equation}\label{eq:diffusion_coefficient}
    D(E) \approx \frac{c l_{\rm coh}}{3} \left[ 4 \left(\frac{E}{E_c}\right)^2 + a_I \left(\frac{E}{E_c}\right) + a_L \left(\frac{E}{E_c}\right)^{1/3} \right],
\end{equation}
where $a_I=0.9$ and $a_L=0.23$ for a Kolmogorov spectrum with $m = 5/3$~\citep{Harari:2013pea}.
The critical energy $E_c$ is given by
\begin{equation}
    E_c \approx Z e B l_{\rm coh} \simeq 9.6 Z \frac{B}{10^{-4}\rm~G} \frac{l_{\rm coh}}{\rm 0.1~kpc}\rm~EeV,
\end{equation}
where $l_{\rm coh} = 0.1\rm~kpc$ is the coherence length.
Note the escape time scale should be larger than the light crossing time scale in order to avoid the superluminal escape.
For charged particles, we use the effective confinement time $t_{\rm conf}={\rm max} [t_{\rm diff}, t_{\rm lc}]$,
while for neutral particles, the escape time scale is equivalent to $t_{\rm lc} = l_b / c$, where $t_{\rm lc}$ represents the light crossing time scale.

We also show the Bethe-Heitler pair production energy loss rate, photomeson production energy loss rate, and photodisintegration energy loss rate for UHECR oxygen and iron nuclei, respectively.
The fate of CR nuclei inside the kiloparsec-scale jet can be characterized by the effective optical depth, $f_{\rm dis} \approx t_{\rm esc} / t_{\rm dis}$, given that the photonuclear reaction is dominated by photodisintegration process. 
As indicated in Fig.~\ref{fig:result-timescale}, both oxygen and iron nuclei could survive up to the highest energy.
We calculate the de-excitation energy loss rate for UHECR oxygen and iron nuclei, respectively.
The ratio between the effective optical depth of the de-excitation process and the Bethe-Heitler pair production process above energies of nuclei interacting with photons at $\varepsilon_b$ is~\citep[e.g.,][]{murase_very-high-energy_2010,aharonian_limitations_2010}
\begin{equation}
    f_{\rm deex} / f_{\rm BH} \sim 1 (0.1)^{\alpha -1} \left(\frac{Z}{8}\right)^{-2} \left(\frac{A}{16}\right)^{0.21\alpha + 1},
\end{equation}
This is consistent with the numerical values displayed in Figure~\ref{fig:result-timescale} considering $\alpha \sim 1$, where $f_{\rm deex} / f_{\rm BH} \sim 1$ for oxygen nuclei and $f_{\rm deex} / f_{\rm BH} \sim 0.4$ for iron nuclei.
Our results reveal the feasibility of detecting de-excitation $\gamma$-rays emitted by light and intermediate-mass nuclei groups, despite the low-energy loss efficiency. 

By examining the blue dotted-dash line, which signifies the two-photon annihilation time scale, and comparing it to the light crossing time scale, as shown in Equation~\ref{eq:opticaldepth-gg}, it is apparent that $\gamma$-rays with high energy levels of up to approximately 100 TeV may be able to flee from the source.

\subsection{Results}
\begin{figure}
	\includegraphics[width=0.45\textwidth]{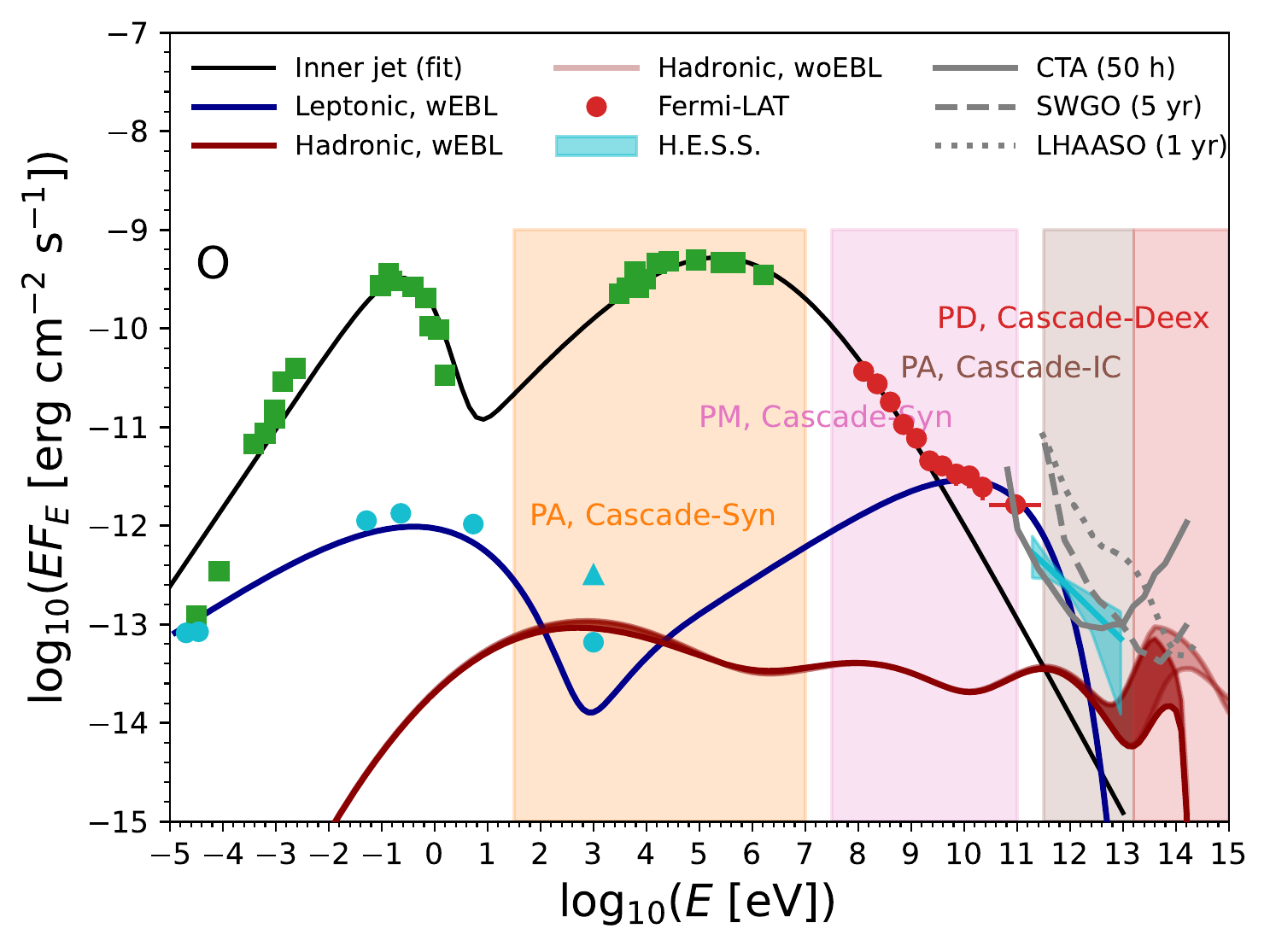}
	\includegraphics[width=0.45\textwidth]{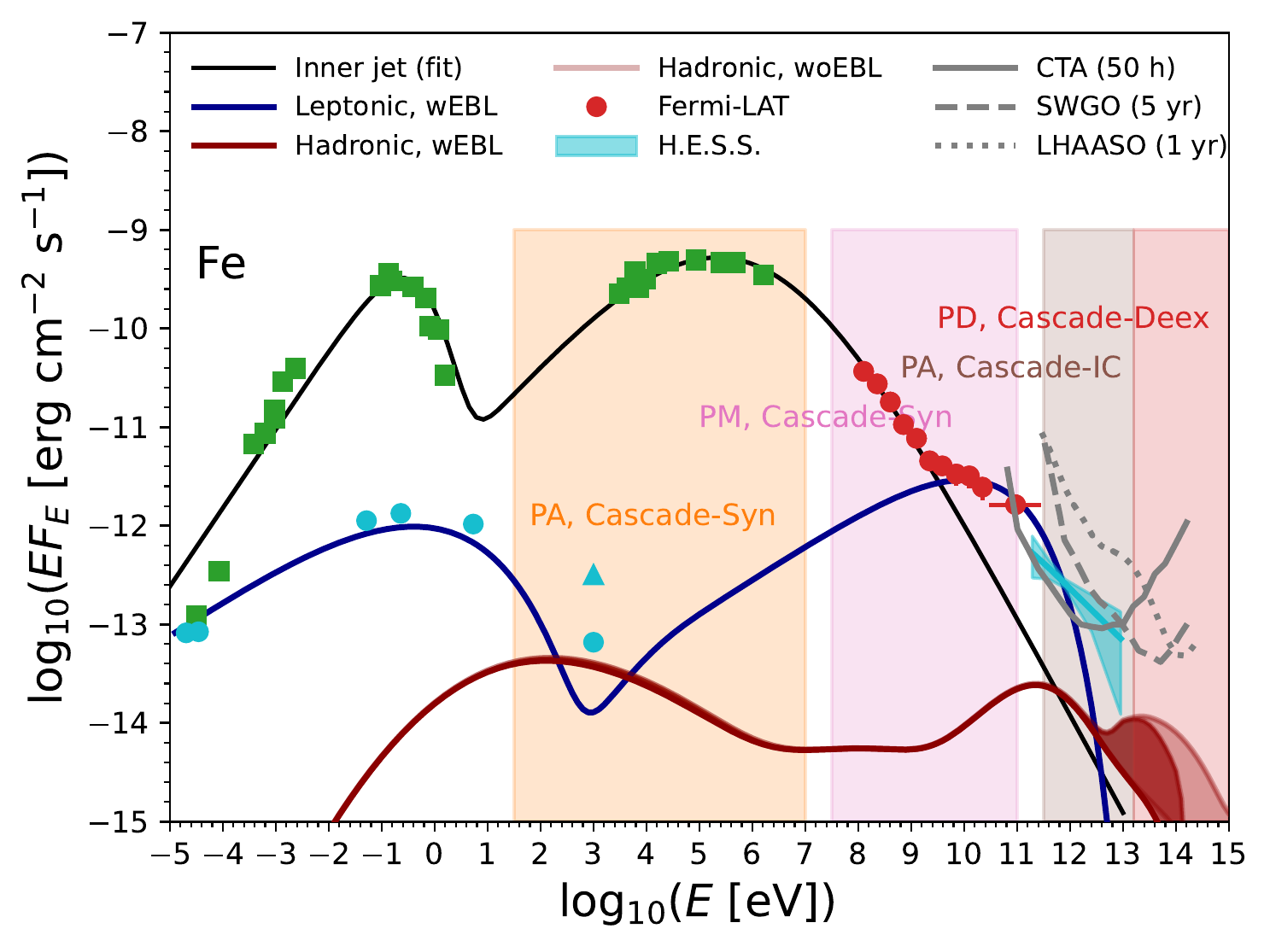}
 \caption{Predicted multi-wavelength SED from the kiloparsec-scale jet after the injection of primary high-energy electrons (solid blue lines) and UHECR nuclei (solid red lines). The solid black lines represent the inner core's SED. The upper panel displays the results from the injection of UHECR oxygen nuclei, while the lower panel shows the results from the injection of UHECR iron nuclei. 
 The green squares and red circles represent data points from the inner core obtained from~\citep{HESS:2018cxr} and cover a range from low-energy radio to optical and high-energy X-ray to GeV bands.  
 The cyan circles represent observed data points from the radio to X-ray band, which correspond to the total flux in the inner region of the large-scale jet and were taken from~\citep{hardcastle_infrared_2006}. The cyan triangles represent the total X-ray flux, which includes an even larger region of the large-scale jet, as reported by~\citep{kataoka_x-ray_2006}. 
 The cyan `butterfly' represents the observed VHE $\gamma$-ray flux by H.E.S.S.~\citep{collaboration_resolving_2020}.
 In addition, this figure displays the expected VHE $\gamma$-ray sensitivities of CTA (50 hours, grey solid lines)~\citep{CTAConsortium:2017dvg}, LHAASO (1 yr, grey dashed lines)~\citep{LHAASO:2019qtb}, and SWGO (5 yr, grey dotted lines)~\citep{Albert:2019afb}.
 }
    \label{fig:result-hadronic}
\end{figure}

\begin{table}
    \centering
	\caption{Physical parameters in the comoving frame used in the leptohadronic model for the kiloparsec-scale jet.}
    \begin{tabular}{l*{1}{c}r}
    Parameter  & Value \\
    \hline
    Viewing angle ($\theta_{\rm ob} [{\rm deg}]$) & 40 \\
    Lorentz factor ($\Gamma_{\rm kpc}$) & 1.05 \\
    Radius ($l_b$ [kpc])  & 0.5  \\
    Magnetic field strength ($B$ [G])  & $1\times 10^{-4}$   \\
    Primary electron index  & 2.3  \\
    Primary electron minimum energy ($\varepsilon_{e, \rm min}$ [eV])  & $2.5\times 10^7$  \\
    Primary electron maximum energy ($\varepsilon_{e, \rm max}$ [eV])  & $2 \times 10^{12} $ \\
    Electron injection energy density ($u_{{\rm inj}, e} [\rm eV~s^{-1}$] & $3$ \\
    \\
    Oxygen minimum energy ($\varepsilon_{\rm O, min}$ [eV]) & $1 \times 10^{17} $\\
    Oxygen maximum energy ($\varepsilon_{\rm O, max}$ [eV]) & $4 \times 10^{18}$ \\
    Oxygen spectral index ($s_{\rm acc}$) & 2.6 \\
    Oxygen injection energy density ($u_{\rm inj, O} [\rm eV~cm^{-3}$] & $300$ \\
    \\
    Iron minimum energy ($\varepsilon_{\rm Fe, min}$ [eV]) &  $1 \times 10^{17} $\\
    Iron maximum energy ($\varepsilon_{\rm Fe, max}$ [eV]) &  $13 \times 10^{18}$ \\
    Iron spectral index ($s_{\rm acc}$) & 2.6 \\
    Iron injection energy density ($u_{\rm inj, Fe} [\rm eV~cm^{-3}$] & $100$ \\
    \end{tabular}
    \label{tab:parameter}
\end{table}
In Fig.~\ref{fig:result-hadronic}, we show the results from the injection of UHECR oxygen and iron nuclei, respectively.
The relevant physical parameters are summarized in Table~\ref{tab:parameter}.
We note that both oxygen and iron nuclei require a minimum injection energy of $10^{17}\rm~eV$.  In this study, we assume that the energy spectrum of CRs accelerated in the kiloparsec-scale jet undergoes a break at $\sim 10^{17}\rm~eV$, where the spectral index $s_{\rm acc}$ may be harder below this energy compared to the value of $2.6$ adopted in this work. This change in the spectral index could be attributed to different acceleration mechanisms. 
In this study,
we assume that UHECRs are more likely to be accelerated via the shear acceleration mechanism, while the low-energy CRs are predominantly contributed by
the diffusive shock acceleration or stochastic acceleration mechanism.
Introducing a low-energy cutoff at $\sim 10^{17}\rm~eV$ is crucial to ensure that the total CR energy does not exceed the jet power.

The solid black lines represent the inner core's SED fitted with Eq.~\ref{eq:two-bump}, while the solid blue and red lines represent the results of the injection of primary electrons and UHECR nuclei, respectively.
We can see that there are four distinct peaks on the predicted multi-wavelength SED from the injection of UHECR nuclei. 
These peaks will be thoroughly discussed in the following paragraphs.
\begin{enumerate}
\item The first peak observed in the keV energy range is attributed to the synchrotron emission generated by electron-positron pairs produced through the Bethe-Heitler pair production process. 
The peak frequency of the synchrotron emission can be roughly estimated using the formula $E_{\rm syn, pk} \approx \delta_{D,\rm kpc} \gamma_e^2 \hbar e B / (m_e c) \simeq 1.1 \times 10^2 \delta_{D,\rm kpc} (B / 10^{-4}{\rm~G}) (\gamma_e / 10^7)^2 \rm~eV$ for electrons with typical energy $E_e \sim 5\delta_{D,\rm kpc}\rm~TeV$. 
\item The second peak observed in the GeV energy range is dominated by the synchrotron emission from electron-positron pairs produced through the photomeson production process. 
These electrons have higher energies, leading to a peak synchrotron frequency estimated to be 
$E_{\rm syn, pk} \simeq 0.1 \delta_{D,\rm kpc} (B / 10^{-4}{\rm~G}) (\gamma_e / 10^{10})^2 \rm~GeV$, assuming typical electron energies of around $E_e \sim 5\delta_{D,\rm kpc}\rm~PeV$. 
As seen in Fig.~\ref{fig:result-hadronic}, the leptonic component dominates the flux observed in the GeV band, compared to the hadronic component.
\item The third peak, located in the TeV energy range, is dominated by the IC emission via electron-positron pairs produced from the Bethe-Heitler pair production process. 
The typical energy of IC scattered photons can be estimated to be $E_{\rm IC, pk} \sim \delta_{D,\rm kpc} \gamma_e^2 \varepsilon \simeq 10 \delta_{D,\rm kpc} (\gamma_e / 10^{7})^2 (\varepsilon / 0.1{\rm~eV}) \rm~TeV$, where $\varepsilon$ is the typical energy of the seed photons in the comoving frame. 
As seen in the figure, the hadronic component can contribute to, and in some cases even dominate, the observed VHE $\gamma$-ray flux above $\sim1~\rm~TeV$. 
\item The last peak in the energy range $\sim 100\rm~TeV$ is predominantly caused by the de-excitation $\gamma$-rays produced during the photodisintegration process of UHECR nuclei. 
These de-excitation $\gamma$-rays can be estimated to have an energy of $E_{\gamma, \rm deex} \approx \delta_{D,\rm kpc} \gamma_A \bar{\varepsilon}_{\gamma, \rm deex} \sim 200 \delta_{D,\rm kpc} (\gamma_A / 10^8) (\bar{\varepsilon}_{\gamma, \rm deex} / 2\rm~MeV)\rm~TeV$, where $\gamma_A = 10^8$ is the Lorentz factor of nuclei.
The uncertainty in the flux of de-excitation $\gamma$-rays, as discussed in Sec.~\ref{sec:fate}, is represented by the shaded area between the two curves at energies beyond $\sim 10\rm~TeV$.
The light-shaded region represents the spectrum without extra-background light (EBL) absorption, while the dark-shaded region accounts for absorption by EBL~\citep{gilmore_semi-analytic_2012}.
Our results indicate that despite strong absorption by the EBL, the detection of de-excitation $\gamma$-rays by current and future ground-based VHE $\gamma$-ray telescopes, such as Cherenkov Telescope Array (CTA)~\citep{CTAConsortium:2017dvg}, 
the Southern Wide-field $\gamma$-ray Observatory (SWGO)~\citep{Albert:2019afb}, and Large High Altitude Air Shower Observatory (LHAASO)~\citep{LHAASO:2019qtb}, is possible. 
Note that LHAASO is included just for comparison because Cen A is not located in its field of view.
\end{enumerate}

\section{Discussion and implications}
\label{sec:dis}
\subsection{Impact of target photon fields}
Note that the detectability of the de-excitation $\gamma$-rays is sensitive to both the target photon energy density and magnetic field strength.
The observed flux in the X-ray band, dominated by the synchrotron emission from pair-induced electrons, increases with a stronger magnetic field.
However, suppose the magnetic field energy density is much lower than the target photon energy density. In that case, the IC emission from pair-induced electrons can surpass the de-excitation $\gamma$-ray flux near its peak energy.
Similarly, a low target photon energy density reduces the energy loss efficiency of the photodisintegration process.
When the target photon field is dense, the escape of de-excitation $\gamma$-rays becomes impossible due to two-photon annihilation.
According to Eq.~\ref{eq:opticaldepth-dis} and Eq.~\ref{eq:opticaldepth-gg}, the ratio between the optical depth of the two-photon annihilation process and the photodisintegration process is 
\begin{equation}
    \tau_{\gamma \gamma} / \tau_{\rm dis} \sim 2 \left(\frac{V}{0.5 c}\right) 2.7^{\alpha - 1} \left(\frac{A}{16}\right)^{0.79\alpha -2} \left(\frac{E_\gamma}{E_A}\right)^{\alpha-1}.
\end{equation}
We can see $\tau_{\gamma \gamma} / \tau_{\rm dis} \sim 2$ for $\alpha \sim 1$.
As pointed out by \cite{Murase:2008mr}, the sources where UHECR nuclei can survive would be optically thin to high-energy $\gamma$-rays~\citep[e.g.,][]{Murase:2008mr}.
In conclusion, the parameter space allowing for the detection of de-excitation $\gamma$-rays is a rather limited, and further study of the available parameter space is necessary.

In our calculations, we initially assume an isotropic distribution of target photons in the comoving frame of the emission region. 
However, we now consider the impact of the anisotropically distributed beamed target photon fields on the observed IC spectrum~\citep[e.g.,][]{bednarek_gevtev_2019, bednarek_morphology_2020}.
When viewing an approaching jet, the IC radiation is reduced compared to the isotropic case due to the small scattering angle between the line-of-sight and the target photon beam direction.
Conversely, the large scattering angle between the line-of-sight and the target photon beam direction enhances the IC radiation from the counter jet.
However, the Doppler beaming effect causes the emission from the counter jet to be reduced by a few factors compared to the emission from the approaching jet.
One significant outcome of the anisotropic IC scattering effect is that the radiation flux from Bethe-Heitler electron-positron pairs can be significantly lower than the flux of de-excitation $\gamma$-rays when viewing an approaching jet, making it easier to identify de-excitation $\gamma$-rays. 

To perform our calculations within the framework of the one-zone model, we assume that the photons from the inner core are uniformly distributed in the comoving frame of the kiloparsec-scale jet.
However, future research should be conducted to study the non-uniform distribution of target photons and the diffusion of UHECR nuclei inside the kiloparsec-scale jet, potentially through Monte Carlo simulations. 

\subsection{Implications for UHECRs}
The injection luminosity of UHECR nuclei can be estimated as~\citep[e.g.,][]{Dermer:2012rg},
\begin{align}\label{eq:power}
    L_{\rm inj}^{\rm UHECR} &\approx 2 \Omega_j \beta_{\rm kpc} c R^2 \Gamma_{\rm kpc}^2 u_{\rm inj},
\end{align}
where $\Omega_j \approx \pi l_b^2/R^2$ is the opening solid-angle and $u_{\rm inj}$ is the comoving frame UHECR injection energy density.
The injection luminosity of UHECR nuclei is $L_{\rm inj}^{\rm O} \sim 7.2 \times 10^{43} \rm~erg~s^{-1}$ oxygen nuclei component and $L_{\rm inj}^{\rm Fe} \sim 2.4 \times 10^{43} \rm~erg~s^{-1}$ iron nuclei component, respectively.
The total jet power should be larger than the value estimated in Eq.~\ref{eq:power}, considering additional contributions from thermal particles, low-energy cosmic rays, radiation fields, and magnetic fields.
The mean jet power of Cen A inferred from the observed enthalpy and age of the southern inner blob is $L_{\rm jet} \sim 10^{43}\rm~erg~s^{-1}$~\citep[e.g.,][]{croston_high-energy_2009, wykes_mass_2013}.
It is apparent that the total jet power estimated in this work is larger than the mean jet power of Cen A. However, higher jet power is still allowed if the jet activity of Cen A has been intermittent. An upper limit of the jet power is the Eddington luminosity $L_{\rm Edd} = 4\pi G m_p M_\bullet/\sigma_T \sim 7 \times 10^{45}\rm~erg~s^{-1}$ with a black hole mass $M_\bullet \sim 5.5 \times 10^7 M_\odot$~\citep[e.g.,][]{Neumayer2007, Cappellari:2008db}. 
If we assume that the energy density in thermal particles and low-energy cosmic rays are not far away from magnetic energy density, we find the total jet power is still less than the Eddington luminosity. 

Cen A, along with other radio galaxies, has been proposed as a candidate source of UHECRs detected on Earth~\citep[e.g.,][]{kimura_ultrahigh-energy_2018, eichmann_ultra-high-energy_2018, matthews_fornax_2018, Bell:2021pkk, Taylor:2023qdy}. 
However, recent studies by~\cite{Eichmann:2022ias} indicate that the CR power of Cen A is about an order of magnitude smaller than its jet power, i.e., $\sim 10^{42}\rm~erg~s^{-1}$, which reduces its likelihood as the main source of UHECRs.
Moreover, this value is about two orders of magnitude smaller than the amount of CR nuclei required in this work.
Nevertheless, if a heavy composition of UHECR nuclei is considered, Cen A could still contribute significantly to the observed UHECRs, as heavy nuclei are less likely to violate the strong quadruple anisotropy constraint~\citep{Eichmann:2022ias}.
To account for the intermediate-scale anisotropies observed in UHECRs, it has been suggested that the scattering of UHECRs emitted from Cen A by the local structure may be a contributing factor~\citep[][]{Bell:2021pkk, Taylor:2023qdy}. 

The escaped flux of charged particles from sources depends on the details of magnetic fields.
The confinement time scale of charged particles could be estimated as $t_{\rm conf} \approx {\rm max}[t_{\rm diff}, t_{\rm lc}]$.
Assuming that UHECRs are isotropized in lobes and/or large-scale structures, the luminosity of escaping cosmic rays are
\begin{equation}
L_{\rm esc}^{A} \approx \frac{1}{t_{\rm conf}} \varepsilon_A^2 \frac{dn^A}{d{\varepsilon_A}}\Gamma_{\rm kpc}^2 2 \pi l_b^3,
\end{equation}
where $dn^A/d\varepsilon_A$ is the steady-state differential energy density of charged particles. 
In Fig.~\ref{fig:UHEnu}, we show the predicted fluxes of CRs and neutrinos that have reached Earth after escaping from their sources.
Our calculations, assuming the rectilinear propagation of CRs without energy losses during propagation, indicated that the expected UHECR flux can be below the observed values, as indicated by black lines.
However, the propagation of UHECRs is strongly impacted by the strength of the intergalactic magnetic field, leading to the magnetic horizon effect, which limits the arrival of CRs at the Earth to only the highest energy particles.

Neglecting energy losses during propagation and the effect of Galactic magnetic fields, the observed flux of CRs at Earth can be expressed by
\begin{equation}
     E_A F_{E_A} = \frac{L_{\rm esc}^A}{4\pi d_L^2} \xi(E_A, d_L, t_{\rm act}),
\end{equation}
where $d_L$ is the source luminosity distance and $t_{\rm act}$ is the activity time. 
The emission from Cen A is assumed to have been continuous but with a recent burst of activity, with $t_{\rm act}=1\rm~Myr$, where $t_{\rm act}$ is the burst lifetime.
The enhancement factor $\xi(E_A, d_L, t_{\rm act})$ can be estimated as~\citep[e.g.,][]{Harari:2020yml, Eichmann:2022ias}
\begin{equation}
    \xi(E_A, d_L, t_{\rm act}) \approx \frac{1}{\mathcal{C} (E_A, d_L)} {\rm exp} \left[-\left(\frac{d_L^2}{0.6 l_D (c t_{\rm act} + d_L)}\right)^{0.8}\right],
\end{equation}
where 
\begin{equation}
    \mathcal{C} (E_A, d_L) = \frac{l_D}{3 d_L} \left[1 - {\rm exp}\left(-3\left(\frac{d_L}{l_D}\right) - 3.5 \left(\frac{d_L}{l_D}\right)^2\right) \right],
\end{equation}
$l_D \equiv 3 D / c$ is the diffusion length and D is the diffusion coefficient (see Eq.~\ref{eq:diffusion_coefficient}).
Note $c t_{\rm act} + d_L$ corresponding to the maximum distance traveled by the observed CRs, where the distance for rectilinear propagation is $d_L$.
Our expected UHECR flux is in agreement with the observed results when taking into account the average intergalactic magnetic field strength of $B = 10^{-9}\rm~G$ and the typical coherence length of $l_{\rm coh} = 100\rm~kpc$~\citep[e.g.,][]{Harari:2020yml}. 
Note that the burst lifetime is much smaller than the typical source lifetime of Cen A, $t_{\rm act} \ll t_{\rm source} \sim 100\rm~Myr$~\citep[e.g.,][]{Taylor:2023qdy}.
The typical time delay $t_{\rm delay}$ between the arrival time of UHECRs and photons that are emitted from sources simultaneously is~\citep[e.g.,][]{Miralda-Escude:1996twc, Murase:2008sa}  
\begin{equation}
    t_{\rm delay} (E_A) \sim 10 \left(\frac{E_A/Z}{10^{18}\rm~eV}\right)^{-2} \left(\frac{d_L}{3\rm~Mpc}\right)^2 \left(\frac{B}{10^{-8} \rm~G}\right)^2  \left(\frac{l_{\rm coh}}{0.1\rm~Mpc}\right)\rm~Myr.
\end{equation}
The time-profile spread is comparable to time delay, $\sigma_d(E_A) \sim t_{\rm delay} (E_A)$~\citep[][]{Takami:2011nn}. 

\subsection{Implications for Neutrinos}
\begin{figure}
    \centering
    \includegraphics[width=0.45\textwidth]{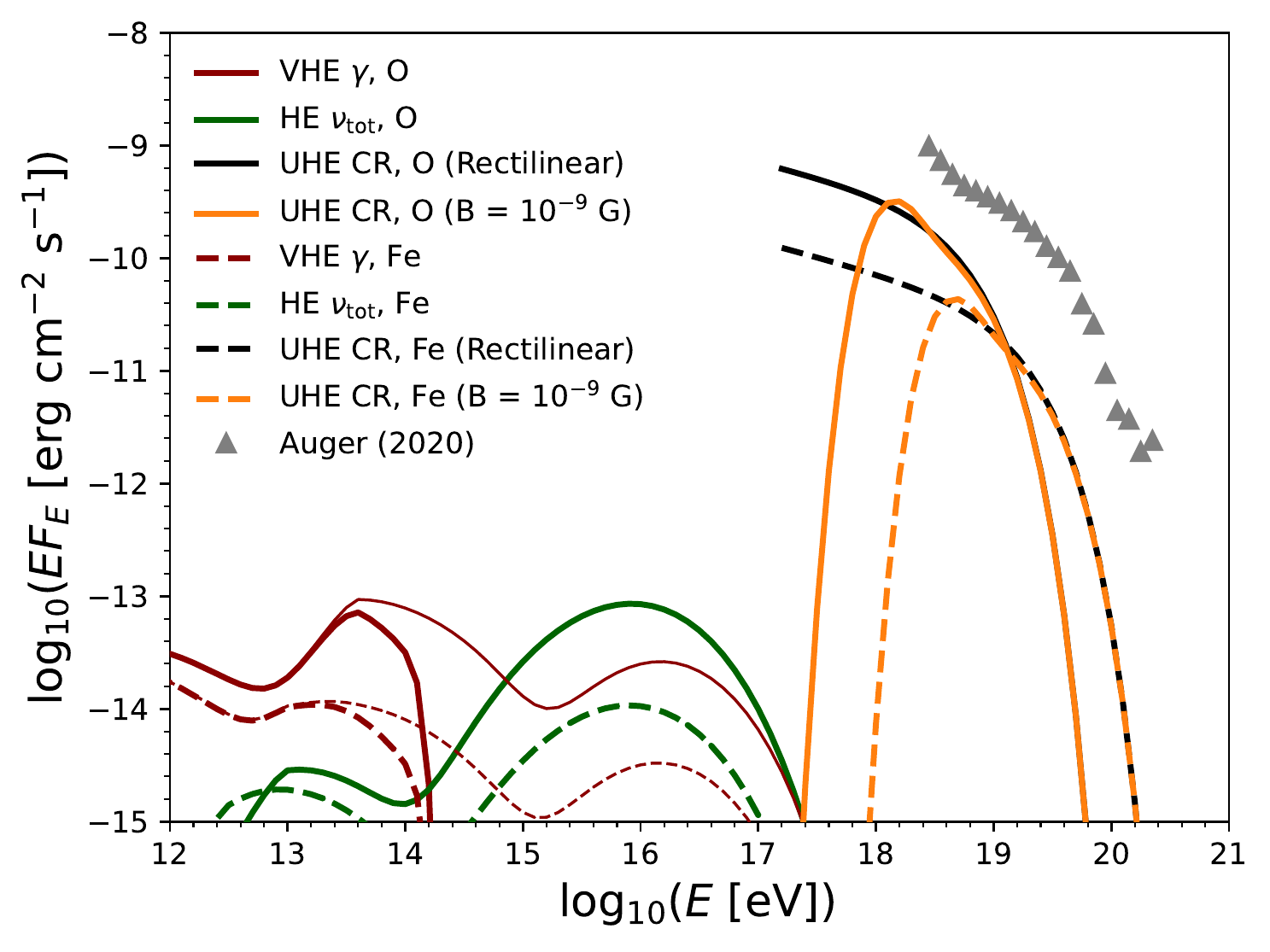}
    \caption{Predicted fluxes of high-energy $\gamma$-rays, neutrinos, and cosmic rays at Earth from Cen A from the kiloparsec-scale jet after the injection of UHECR oxygen nuclei (solid lines) and iron nuclei (dashed lines). For high-energy $\gamma$-rays, we also show the unattenuated spectrum without considering EBL absorption (thin lines).} The grey triangles are the diffuse flux of UHECRs measured by the Pierre Auger Collaboration~\citep{PierreAuger:2020qqz}.
    \label{fig:UHEnu}
\end{figure}

In Fig.~\ref{fig:UHEnu}, we also show the predicted all-flavor energy spectrum of high-energy neutrinos from pion decay and neutron $\beta$-decay during the photodisintegration process.
Although we adopt the numerical approach, the observed all-flavor energy spectrum of high-energy neutrinos produced from the pion decay process can be analytically estimated with the following formula,
\begin{align}
    E_\nu F_{E_\nu}^{(\rm mes)} &\approx \delta_{D,\rm kpc}^4 \frac{1}{4\pi d_L^2} \frac{3}{8} f_{\rm mes} (\varepsilon_A) \varepsilon_A L'_{\varepsilon_A},
\end{align}
where $\varepsilon_A L'_{\varepsilon_A}$ is the comoving frame cosmic-ray injection luminosity, $E_\nu \approx \delta_{D,\rm kpc} 0.05 (\varepsilon_A / A) \sim \delta_{D,\rm kpc} 10^{16} \rm~eV$ is the typical neutrino energy in the observer frame for $\varepsilon_A \sim 3\times 10^{18}\rm~eV$ and $A = 16$.
Note both direct neutrino contribution by the photomeson production process on nuclei and indirect neutrino contribution by the photomeson production process on secondary neutrons and protons are included~\citep{Murase:2010gj, Zhang:2018agl}.
The observed energy spectrum of anti-electron neutrinos from neutron $\beta$-decay can be written as
\begin{align}
    E_\nu F_{E_\nu}^{(\beta_{\rm dec})} &\approx \delta_{D,\rm kpc}^4 \frac{1}{4\pi d_L^2} \kappa_{\beta_{\rm dec}} \frac{t_{\rm esc}^n}{\gamma_n \tau_n} \xi_n f_{\rm dis} (\varepsilon_A) \varepsilon_A L'_{\varepsilon_A},
\end{align}
where $\gamma_n$ is the Lorentz factor of neutrons, $\tau_n \simeq 879.6\rm~s$ is the mean lifetime of free neutrons~\citep{Workman:2022ynf}, $\kappa_{\beta_{\rm dec}} \approx \langle\varepsilon_\nu\rangle / m_n c^2$ is the inelasticity, $m_n$ is the neutron mass in the neutron rest frame, $\xi_n \sim 1/2$ is the fraction of neutrons in the emitted nucleons.
The average electron kinetic energy in the neutron rest frame is $\langle\varepsilon_e\rangle - m_e c^2 \approx 0.30\rm~MeV$.
The typical neutrino energy  in the neutron rest frame can be estimated as $\langle\varepsilon_\nu\rangle \approx Q_\beta - (\langle\varepsilon_e\rangle - m_e c^2) - (\langle\varepsilon_p\rangle - m_p c^2) \simeq 0.48\rm~MeV$, when measured in the neutron rest frame, where $Q_\beta \approx 0.78\rm~MeV$ is the $Q$-value representing the difference between the initial and final mass energies, $\langle\varepsilon_p\rangle - m_p c^2 \sim 0.3\rm~keV$, is the proton recoil kinetic energy. 
The typical neutrino energy in the observer frame is $E_\nu \approx \delta_{D,\rm kpc} \langle\varepsilon_\nu\rangle \gamma_n \sim \delta_{D,\rm kpc} 3\times 10^{13}\rm~eV$ for $\varepsilon_A \sim 1\times 10^{18}\rm~eV$ and $A = 16$.
Note that $\beta$-decay from nuclei is not included in this study.

We can see the neutrino flux lies in the $\sim 10\rm~PeV$ energy range dominated by the pion decay process, while the neutron $\beta$-decay process dominates the neutrino flux in the lower energy range, $\sim 0.01\rm~PeV$, 
High-energy neutrino emission from Cen A can be searched for by the next-generation neutrino telescopes, such as KM3Net~\citep{KM3Net:2016zxf} and IceCube-Gen2~\citep{IceCube-Gen2:2020qha}. 
As Cen A located in the Southern sky, KM3Net is more sensitive than IceCube-Gen2 for detecting high-energy neutrinos from Cen A.
However, the flux of high-energy neutrinos predicted in this work is $\sim 1 - 2$ orders smaller than the prediction of the magnetically-powered corona model for Cen A~\citep[][]{Kheirandish:2021wkm}, and the detection of high-energy neutrinos seem challenging.

\section{Summary}\label{sec:sum}
In this study, we evaluated the feasibility of detecting de-excitation VHE $\gamma$-rays through our numerical code that self-consistently considers both nuclear and electromagnetic cascades.
The accuracy of the code has been validated through comparisons with Monte Carlo simulations using a modified version of {\sc CRPropa 3}. 
Our code can be used to explore the behavior of UHECR nuclei in astrophysical sources under the assumption of a one-zone model.
 
We then applied our numerical code to the closest radio galaxy, Cen A, which we considered a potential accelerator of UHECR nuclei in its kiloparsec-scale jet. 
In our model, the primary target photons are the beamed inner core emission that illuminates the jet when viewed along its axis, but this emission is greatly reduced when viewed from Earth due to the Doppler beaming effect.

Our results, assuming the dominant injection of UHECR nuclei consisting of oxygen and/or iron, indicate that the de-excitation VHE $\gamma$-rays will be the dominant contributor to the multi-wavelength spectrum at $\gtrsim 10 - 100\rm~TeV$ if the UHECR nuclei are dominated by oxygen-group components. The de-excitation VHE $\gamma$-rays from Cen A could be detected by current and future ground-based VHE $\gamma$-ray telescopes.
The results obtained in this work provide valuable insight into the composition of UHECR nuclei in nearby extragalactic sources.

\section*{Acknowledgements}
K.M. acknowledges John Beacom, Charles Dermer, and Asaf Pe'er for early discussions in 2011-2012. The work of B.T.Z. is supported by KAKENHI No.~20H01901. The work of K.M. is supported by the NSF Grant No.~AST-1908689, No.~AST-2108466 and No.~AST-2108467, and KAKENHI No.~20H01901 and No.~20H05852.

%%%%%%%%%%%%%%%%%%%%%%%%%%%%%%%%%%%%%%%%%%%%%%%%%%
\section*{Data Availability}
The data developed for the calculation in this work is available upon request.
The code used will be made public in the future as a part of the AMES. 

\bibliographystyle{mnras}
\bibliography{CenA} % if your bibtex file is called example.bib

%%%%%%%%%%%%%%%%%%%%%%%%%%%%%%%%%%%%%%%%%%%%%%%%%%

%%%%%%%%%%%%%%%%% APPENDICES %%%%%%%%%%%%%%%%%%%%%

\appendix
\section{Numerical method and related physical processes}\label{app:solution}
The coupled transport equation given by Eq.~\ref{eq:transport} can be discretized as
\begin{align}\label{eq:transport-discretize}
    \frac{d}{dt} n_{a, i} &=  -n_{a, i} \mathcal{A}_{a, i}
    + \sum_{j \geqslant i} n_{a, j} \mathcal{B}_{a, j \to i} \nonumber \\
    &+ \sum_b \sum_{j \geqslant i} n_{b, j} \mathcal{C}_{b, j \to a, i} 
    + \dot{n}_{a, i}^{\rm inj},
\end{align}
where the indices $i$ and $j$ represent different energy bins.
The above equations can be numerically solved with the first-order implicit scheme~\citep[e.g.,][]{lee_propagation_1998},
\begin{align}
    \frac{n_{a, i}^{m+1} -n_{a, i}^{m}}{\Delta t} &= -n_{a, i}^{m+1} \mathcal{A}_{a, i}  + \sum_{j \geqslant i} n_{a, j}^{m+1} \mathcal{B}_{a, j \to i} \nonumber \\ &+ \sum_b \sum_{j \geqslant i} n_{b, j}^{m+1} \mathcal{}_{b, j \to a, i} + \dot{n}_{a, i}^{{\rm inj},m}, 
\end{align}
and we have
\begin{align}\label{eq:solve}
    n_{a, i}^{m+1} = \frac{\frac{n_{a, i}^{m}}{\Delta t} + \sum_{j > i} n_{a, j}^{m+1} \mathcal{B}_{a, j \to i} + \sum_b \sum_{j \geqslant i} n_{b, j}^{m+1} \mathcal{C}_{b, j \to a, i} + \dot{n}_{a, i}^{{\rm inj},m}}{\frac{1}{\Delta t} + \mathcal{A}_{a, i} - \mathcal{B}_{a, i \to i}},\,\,\,\,\,\,\,\,
\end{align}
where $\Delta t$ is the time step, the index $m$ represents the current particle number density at time $t$, and the index $m+1$ represents the particle number density at time $t + \Delta t$.
In order to keep the accuracy, we adopt the method used in \cite{M09, M12, MB12}, where at each time step the solution of Eq.~\ref{eq:solve} is found when the number density of each species converges.  

To check the availability of our transport code, which is a part of the {\sc Astrophysical Multimessenger Emission Simulator (AMES)}, we compared our results with Monte Carlo simulations with a modified version of {\sc CRPropa 3}.
For the Monte Carlo simulations, instead of implementing the injection term $\dot{n}_{a}^{\rm inj}$, we choose $\dot{n}_{a}^{\rm inj}= 0$ when solving Eq.~\ref{eq:transport} just for the comparison purpose.
We also neglect the escape term. We consider a spherical blob with a radius of $l_b = 1\rm~pc$ moving toward the observer with a Lorentz factor of $\Gamma = 10$.
The target photon fields in the comoving frame of the blob can be described by Eq.~\ref{eq:target}, which are isotropically distributed inside the blob with $\varepsilon_{b} = 1\rm~eV$, $\alpha_l = 1$, $\alpha_h =2.5$ and $u_{\rm ph} = 10^7\rm~eV$ is the comoving frame target photon energy density.
The injected energy spectrum of oxygen nuclei follows a power-law distribution with an exponential cutoff, where $\varepsilon_{\rm max} = 8 \times 10^{\rm 17}\rm~eV$ and $s_{\rm acc} = 2$.
The dynamical time scale of the system is given by $t_{\rm dyn} = l_b/V$, where $V$ is the characteristic velocity (e.g., shock crossing time during which photons are generated).

\begin{figure}
	\includegraphics[width=\linewidth]{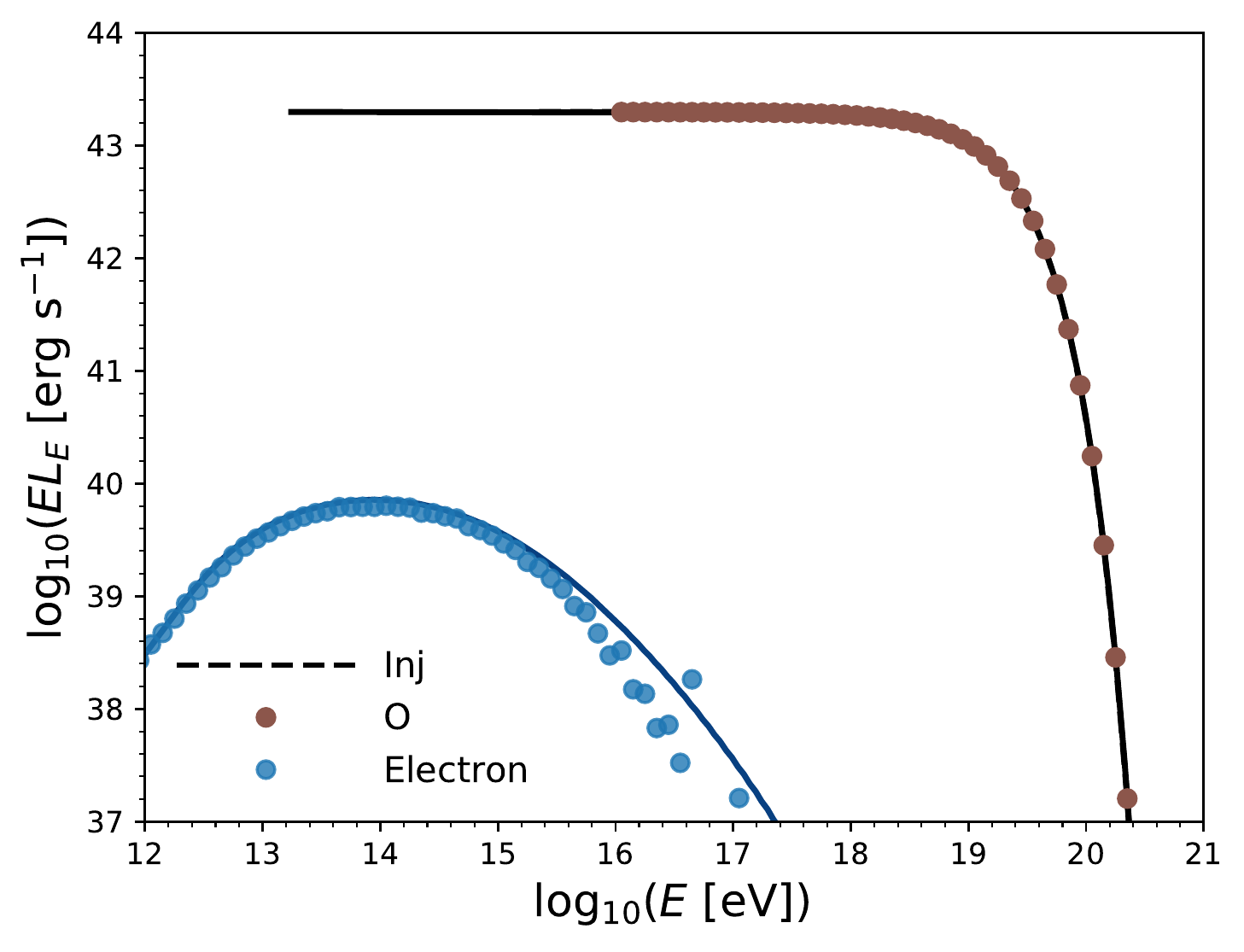}
	\includegraphics[width=\linewidth]{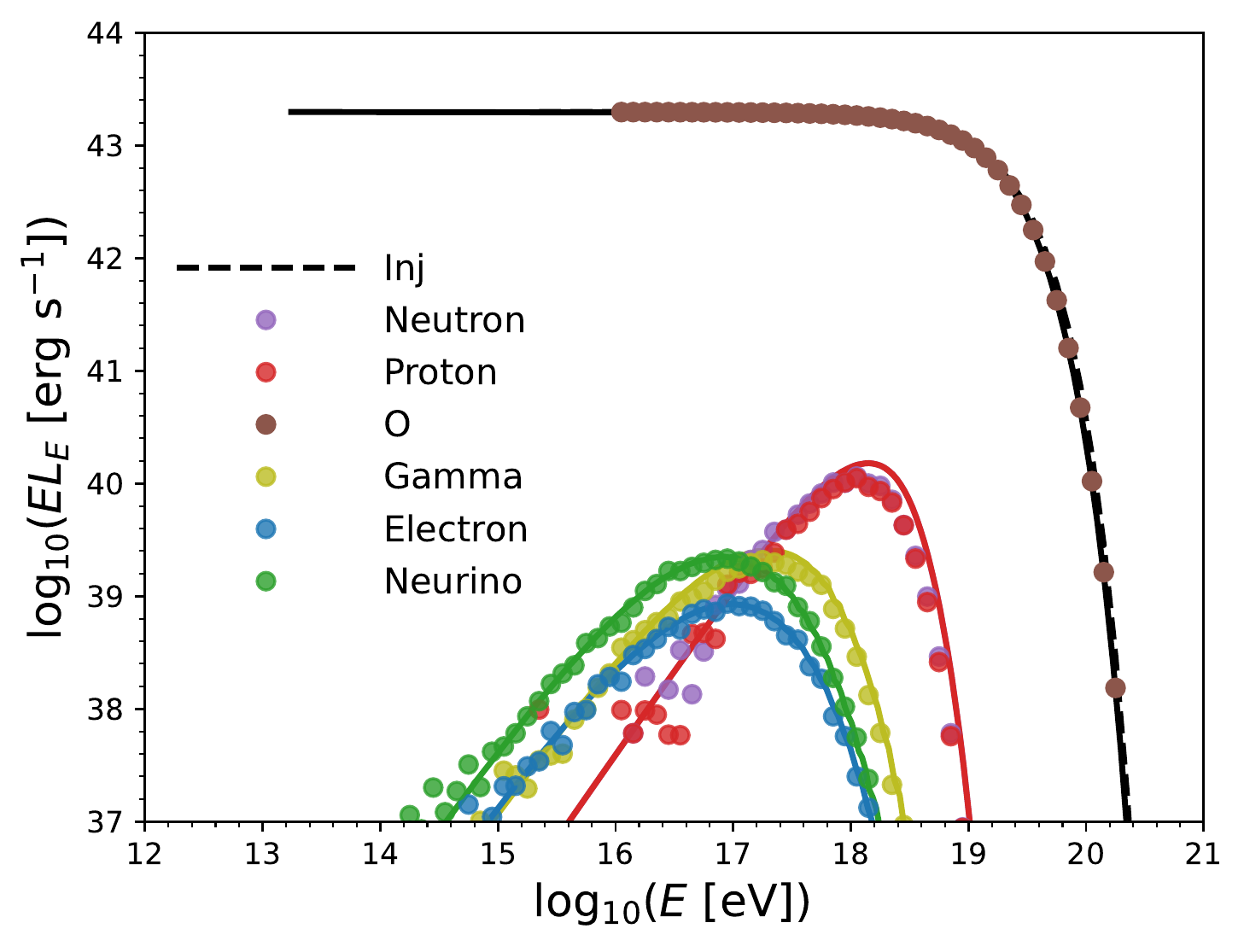}
	\includegraphics[width=\linewidth]{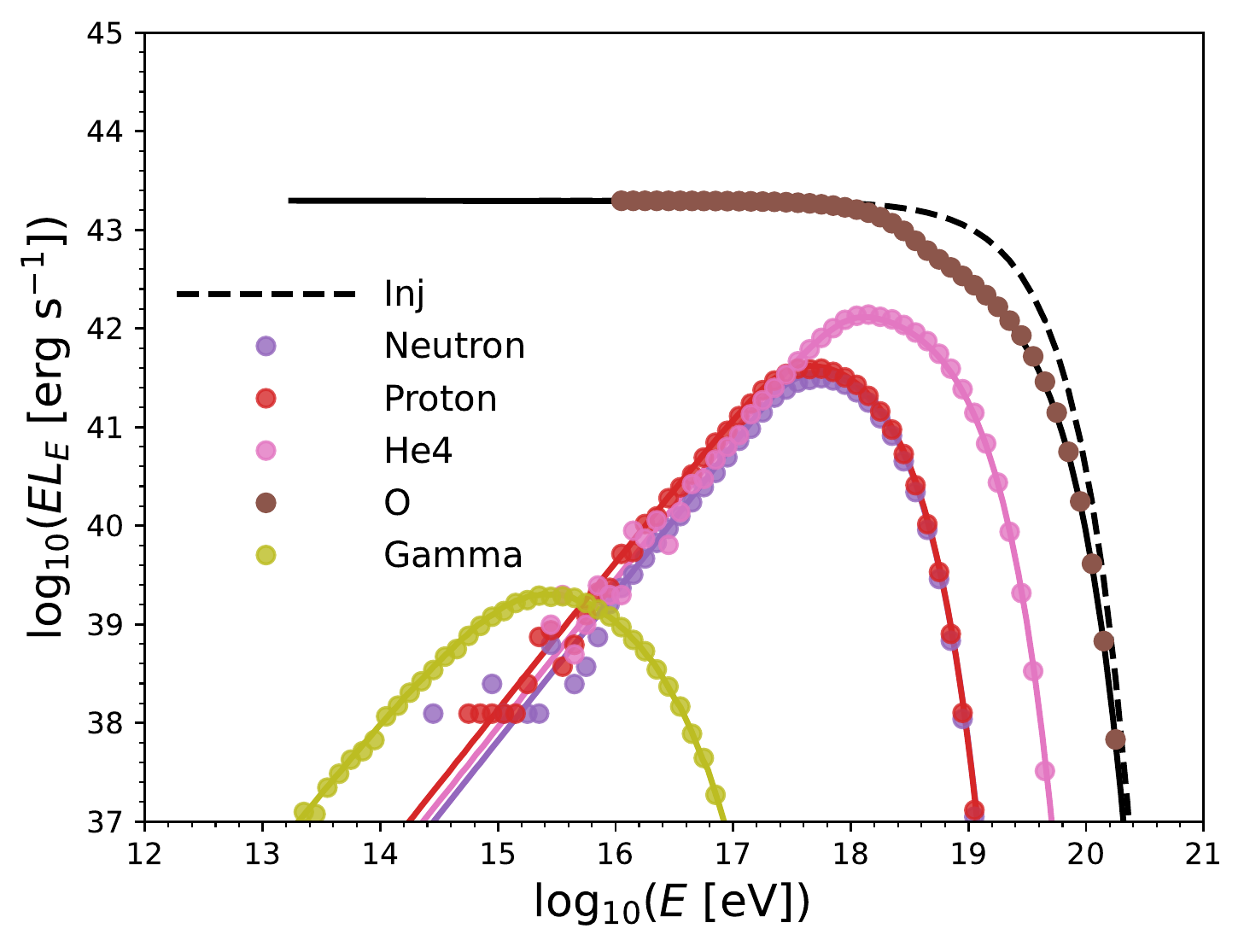}
	\caption{Spectral energy distributions of primary nuclei and secondary particles, including daughter nuclei and other particles, which are calculated with our kinetic code (solid curves) and Monte Carlo code (dots), where we consider the Bethe-Heitler pair production process (upper panel), photomeson production process (middle panel), and photodisintegration process (lower panel), respectively.}\label{app-fig:compare}
\end{figure}

In Fig.~\ref{app-fig:compare}, we compare the output from our kinetic code with what derived from Monte Carlo simulations using a modified version of {\sc CRPropa 3}~\citep[][]{batista_crpropa_2016, zhang_neutral_2020}. 
We compare the spectrum of surviving nuclei and generated photons, electrons, and neutrinos, considering three main hadronic processes including the Bethe-Heitler pair production process (upper panel), the photomeson production process (middle panel), and the photodisintegration process (lower panel).
The results from the two calculation methods are consistent with each other, except for stochastic fluctuations. Indeed, the difference in the high-energy part of Bethe-Heilter pairs is due to the limited sampling of target photons for {\sc CRPropa 3} in handling the Bethe-Heilter process. 

The details of the value of the coefficients $\mathcal{A}$, $\mathcal{B}$, and $\mathcal{C}$ for various particles are discussed below.
Note all the quantities are in the comoving frame.

\subsection{Photon}
The kinetic equation for $\gamma$-rays is~\citep[see also Supplemental Material of][]{murase_new_2018}
\begin{align}\label{eq:transport-photon}
    \frac{\partial n_{\varepsilon_\gamma}^{\gamma}}{\partial t} &= -n_{\varepsilon_\gamma}^{\gamma} \left(\frac{1}{t_{\rm esc}^\gamma} + \frac{1}{t_{\gamma \gamma}}\right) \nonumber \\ &+ \frac{\partial}{\partial t} \left(n_{\varepsilon_\gamma}^{\rm syn} + n_{\varepsilon_\gamma}^{\rm IC} + n_{\varepsilon_\gamma}^{\rm phmes} + n_{\varepsilon_\gamma}^{\rm deex}\right) + \dot{n}_{\varepsilon_\gamma}^{\rm inj},
\end{align}
where $t_{\rm esc}^\gamma$ is the photon escape time scale, $t_{\rm \gamma \gamma}$ is the interaction time scale of high-energy $\gamma$-rays with ambient target photon fields,
$\partial n_{\varepsilon_\gamma}^{\rm syn}/\partial t$ is the $\gamma$-ray generation rate from synchrotron emission process of all the charged particles, $\partial n_{\varepsilon_\gamma}^{\rm IC}/\partial t$ is the $\gamma$-ray generation rate from the inverse-Compton process of electrons, $\partial n_{\varepsilon_\gamma}^{\rm phmes}/\partial t$ is the $\gamma$-ray generation rate from the photomeson production process of both nucleons and nuclei, $\partial n_{\varepsilon_\gamma}^{\rm phdis}/\partial t$ is the $\gamma$-ray generation rate from the photodisintegration process via de-excitation process and $\dot{n}_{\varepsilon_\gamma}^{\rm inj}$ is the primary $\gamma$-ray injection rate (e.g., from external regions).

Based on Eq.~\ref{eq:transport-photon}, the coefficient $\mathcal{A}$ in Eq.~\ref{eq:transport} corresponds to
\begin{equation}
    \mathcal{A}_\gamma = \frac{1}{t_{\rm esc}^\gamma} + \frac{1}{t_{\rm \gamma \gamma}}.
\end{equation}
The coefficient $\mathcal{B}$ is set to be zero
\begin{equation}
    \mathcal{B}_{\gamma \to \gamma} = 0,
\end{equation}
which means that incident high-energy $\gamma$-rays disappear once they annihilate with target photons.
The coefficient $\mathcal{C}$ is
\begin{equation}
    \mathcal{C}_\gamma = \frac{\partial}{\partial t} \left(n_{\varepsilon_\gamma}^{\rm syn} + n_{\varepsilon_\gamma}^{\rm IC} + n_{\varepsilon_\gamma}^{\rm phmes} + n_{\varepsilon_\gamma}^{\rm phdis}\right).
\end{equation}
The escape time scale for photons is set to $t_{\rm esc}^\gamma = t_{\rm lc} = R / c$, where $t_{\rm lc}$ is the light crossing time scale and $R$ is the radius of the emission region.

\subsection{Electron}
The kinetic equation for high-energy electrons (and positrons) is
\begin{align}\label{eq:transport-electron}
    \frac{\partial n_{\varepsilon_e}^{e}}{\partial t} &= -\frac{n_{\varepsilon_e}^{e}}{t_{\rm esc}^e} 
    - \frac{\partial}{\partial \varepsilon_e} \left[(P_{\rm ad} + P_{\rm syn}^e + P_{\rm IC}^e) n_{\varepsilon_e}^{e} \right] \nonumber \\ &+ \frac{\partial}{\partial t} \left(n_{\varepsilon_e}^{\gamma \gamma} + n_{\varepsilon_e}^{\rm BH} + n_{\varepsilon_e}^{\rm phmes} + n_{\varepsilon_e}^{\beta_{\rm dec}}\right) + \dot{n}_{\varepsilon_e}^{\rm inj},
\end{align}
where $t_{\rm esc}^e$ is the electron escape time scale, $P_{\rm ad}$ is the adiabatic energy loss rate, $P_{\rm syn}^e$ is the synchrotron energy loss rate, $P_{\rm IC}^e$ is the inverse-Compton energy loss rate, $\partial n_{\varepsilon_e}^{\rm BH} / \partial t$ is the electron-positron generation rate from the Bethe-Heitler pair production process, $\partial n_{\varepsilon_e}^{\rm phmes} / \partial t$ is the electron-positron generation rate from the photomeson production process, $n_{\varepsilon_e}^{\beta_{\rm dec}}$ is the electron generation rate from $\beta$ decay, $\dot{n}_{\varepsilon_e}^{\rm inj}$ is the primary electron injection rate.

For electromagnetic cascades inside the source, we adopt the continuous energy-loss approximation, as in \citet{Murase:2014bfa}, where the second term in the right hand of Eq.~\ref{eq:transport-electron} can be rewritten as $\frac{\partial}{\partial \varepsilon_e} \left[(P_{\rm ad} + P_{\rm syn}^e + P_{\rm IC}^e) n_{\varepsilon_e}^{e} \right] = \frac{\partial}{\partial \varepsilon_e} \left[P_{\rm ad} + P_{\rm syn}^e + P_{\rm IC}^e \right]  n_{\varepsilon_e}^{e} + \frac{\partial}{\partial \varepsilon_e} \left[n_{\varepsilon_e}^{e} \right] (P_{\rm ad} + P_{\rm syn}^e + P_{\rm IC}^e)$.
We can express the coefficient $\mathcal{A}$ as
\begin{equation}
    \mathcal{A}_e = \frac{1}{t_{\rm esc}^e} + \frac{\partial}{\partial \varepsilon_e} \left[P_{\rm ad} + P_{\rm syn}^e + P_{\rm IC}^e \right],
\end{equation}
where the discretization formula is
\begin{equation}
    \mathcal{A}_{e, i} = \frac{1}{t_{\rm esc}^e}\bigg|_{\varepsilon = \varepsilon_i} + \frac{1}{\varepsilon_{i+1/2} - \varepsilon_{i-1/2}}\left(P_{\rm ad} + P_{\rm syn}^e + P_{\rm IC}^e\right)|_{\varepsilon = \varepsilon_i}.
\end{equation}
The coefficient $\mathcal{B}$ can be written as
\begin{equation}
    \mathcal{B}_{e, j \to e, i} = \delta_{j}^{i+1} \frac{1}{\varepsilon_{j+1/2} - \varepsilon_{j-1/2}}\left(P_{\rm ad} + P_{\rm syn}^e + P_{\rm IC}^e\right)|_{\varepsilon = \varepsilon_j},
\end{equation}
where $\delta$ is the Kronecker delta function, the index $i$ and $j$ represent electron energy index~\citep[e.g.,][]{MB12, kalashev_simulations_2015}.
The coefficient $\mathcal{C}$ can be written as
\begin{equation}
    \mathcal{C}_e = \frac{\partial}{\partial t} \left(n_{\varepsilon_\gamma}^{\gamma \gamma} + n_{\varepsilon_\gamma}^{\rm BH} + n_{\varepsilon_\gamma}^{\rm phmes} + n_{\varepsilon_\gamma}^{\beta_{\rm dec}}\right).
\end{equation}

In general, the escape time depends on the diffusion coefficient or details of magnetic fields. In the limit that the magnetic confinement is sufficiently long, which is valid for electrons in the energy range of interest, one may approximate the escape time scale by 
\begin{equation}
t_{\rm esc}^e \approx t_{\rm adv} = \frac{l_b}{V},
\end{equation}
where $t_{\rm adv}$ is the advection time scale and $V$ is the expansion speed of the emission region.
The adiabatic energy loss time scale can be estimated,
\begin{equation}
t_{\rm ad}^{-1} \equiv \frac{1}{\varepsilon_e} \frac{d\varepsilon_e}{dt} \approx \left(\frac{l_b}{V_{\rm adv}}\right)^{-1},
\end{equation}
where the adiabatic energy loss rate is $P_{\rm ad} = \varepsilon_e t_{\rm ad}^{-1}$, $l_b$ is the comoving size of the blob, and $V_{\rm adv} \sim V$ is the advection velocity at the dissipation region.

\subsection{Neutrino}
The kinetic equation for neutrinos is
\begin{align}
    \label{eq:transport-neutrino}
    \frac{\partial n_{\varepsilon_\nu}^{\nu}}{\partial t} &= -\frac{n_{\varepsilon_\nu}^{\nu}}{t_{\rm esc}} + \frac{\partial}{\partial t} \left(n_{\varepsilon_\nu}^{\rm phmes} + n_{\varepsilon_\nu}^{\beta_{\rm dec}}\right),
    %+ \dot{n}_{\varepsilon_\nu}^{\rm inj},
\end{align}
where $t_{\rm esc}^\nu= t_{\rm lc}$ is the neutrino escape time scale, $\partial n_{\varepsilon_\nu}^{\rm phmes} / \partial t$ is the neutrino generation rate from the photomeson production process, and $\partial n_{\varepsilon_\nu}^{\beta_{\rm dec}} / \partial t$ is the (anti-)electron neutrino generation rate from $\beta$ decay.
%, $\dot{n}_{\varepsilon_\nu}^{\rm inj}$ is the primary neutrino injection rate.

The coefficient $\mathcal{A}$ is
\begin{equation}
    \mathcal{A}_\nu = \frac{1}{t_{\rm esc}^\nu}.
\end{equation}
The coefficient $\mathcal{B}$ is
\begin{equation}
    \mathcal{B}_\nu = 0.
\end{equation}
The coefficient $\mathcal{C}$ is 
\begin{equation}
    \mathcal{C}_\nu = \frac{\partial}{\partial t} \left(n_{\varepsilon_\nu}^{\rm phmes} + n_{\varepsilon_\nu}^{\beta_{\rm dec}}\right).
\end{equation}

\subsection{Neutron}
The kinetic equation for neutrons is
\begin{align}\label{eq:transport-neutron}
    \frac{\partial n_{\varepsilon_n}^{n}}{\partial t} = -\frac{n_{\varepsilon_n}^{n}}{t_{\rm esc}^n} -\frac{n_{\varepsilon_n}^{n}}{t_{\rm phmes}}-\frac{n_{\varepsilon_n}^{n}}{t_{\beta_{\rm dec}}} + \frac{\partial}{\partial t} \left(n_{\varepsilon_n}^{\rm phmes} + n_{\varepsilon_n}^{\rm phdis}\right),
    %+ \dot{n}_{\varepsilon_n}^{\rm inj},
\end{align}
where $t_{\rm esc}^n= t_{\rm lc}$ is the neutron escape time scale, $t_{\rm phmes}$ is the photomeson production time scale, $t_{\beta_{\rm dec}}$ is the neutron lifetime, $\partial n_{\varepsilon_n}^{\rm phmes} / \partial t$ is the neutron generation rate from the photomeson production process, and $\partial n_{\varepsilon_n}^{\rm phdis} / \partial t$ is the neutron generation rate from nuclear photodisintegration.

The coefficient $\mathcal{A}$ for neutron is
\begin{equation}
    \mathcal{A}_n = \frac{1}{t_{\rm esc}^n} + \frac{1}{t_{\rm phmes}} + \frac{1}{t_{\beta_{\rm dec}}}.
\end{equation}
The coefficient $\mathcal{B}$ is,
\begin{equation}
    \mathcal{B}_n = \frac{\partial n_{\varepsilon_n}^{\rm phmes}}{\partial t},
\end{equation}
where $\partial n_{\varepsilon_n}^{\rm phmes}/\partial t$ is the generation rate of the neutron from the photomeson production process when the primary particle is a neutron.
The coefficient $\mathcal{C}$ is
\begin{equation}
    \mathcal{C}_n = \frac{\partial}{\partial t} \left(n_{\varepsilon_n}^{\rm phmes} + n_{\varepsilon_n}^{\rm phdis}\right),
\end{equation}
where $\partial n_{\varepsilon_n}^{\rm phmes}/\partial t$ is the generation rate of neutrons from the photomeson production process when the primary particle is a proton or a nucleus.

\subsection{Proton}
The kinetic equation for protons is
\begin{align}\label{eq:transport-proton}
    \frac{\partial n_{\varepsilon_p}^{p}}{\partial t} &= -\frac{n_{\varepsilon_p}^{p}}{t_{\rm esc}} -\frac{n_{\varepsilon_p}^{p}}{t_{\rm phmes}} + \frac{\partial}{\partial \varepsilon_p} \left[(P_{\rm ad} + P_{\rm BH}^p + P_{\rm syn}^p) n_{\varepsilon_p}^{p} \right] \nonumber \\ &+ \frac{\partial}{\partial t} \left(n_{\varepsilon_p}^{\rm phmes} + n_{\varepsilon_p}^{\beta_{\rm dec}} + n_{\varepsilon_p}^{\rm phdis}\right) + \dot{n}_{\varepsilon_p}^{\rm inj},
\end{align}
where $t_{\rm esc}^p$ is the proton escape time scale, $t_{\rm phmes}$ is the photomeson production interaction time scale, $P_{\rm BH}$ is the energy loss rate due to the Bethe-Heitler pair production process, $P_{\rm syn}^p$ is the proton synchrotron energy loss rate, $\partial n_{\varepsilon_p}^{\rm phmes} / \partial t$ is the proton generation rate from the photomeson production process, $n_{\varepsilon_p}^{\beta_{\rm dec}}$ is the proton generation rate from $\beta$ decay, $\partial n_{\varepsilon_p}^{\rm phdis} / \partial t$ is the proton generation rate from nuclear photodisintegration, $\dot{n}_{\varepsilon_p}^{\rm inj}$ is the primary proton injection rate.

For the Bethe-Heitler pair production process, we adopt the continuous energy loss approximation.
The efficient $\mathcal{A}$ is
\begin{equation}
    \mathcal{A}_p = \frac{1}{t_{\rm esc}^p} + \frac{1}{t_{\rm phmes}} + \frac{\partial}{\partial \varepsilon_p} (P_{\rm ad} + P_{\rm BH}^p + P_{\rm syn}^p),
\end{equation}
where the discretization form is 
\begin{equation}
    \mathcal{A}_{p, i} = \frac{1}{t_{\rm esc}^p}\bigg|_i + \frac{1}{t_{\rm phmes}}\bigg|_i + \frac{1}{\varepsilon_{i+1/2} - \varepsilon_{i-1/2}}(P_{\rm ad} + P_{\rm BH}^p + P_{\rm syn}^p)|_{\varepsilon = \varepsilon_i}.
\end{equation}
The coefficient $\mathcal{B}$ is
\begin{equation}
    \mathcal{B}_{p, j \to p, i} = \frac{\partial n_{\varepsilon_p}^{\rm phmes}}{\partial t} + \delta_{j}^{i+1} \frac{1}{\varepsilon_{j+1/2} - \varepsilon_{j-1/2}}(P_{\rm ad} + P_{\rm BH}^p + P_{\rm syn}^p)|_{\varepsilon = \varepsilon_j},
\end{equation}
where $\partial n_{\varepsilon_p}^{\rm phmes} / \partial t$ is the generation rate of protons from the photomeson production process when the primary particle is a proton.
The coefficient $\mathcal{C}$ is written as
\begin{equation}
    \mathcal{C}_p = \frac{\partial}{\partial t} \left(n_{\varepsilon_p}^{\beta_{\rm dec}} + n_{\varepsilon_p}^{\rm phmes} + n_{\varepsilon_p}^{\rm phdis}\right).
\end{equation}
For the escape processes, we consider both diffusion and advection. 
The confinement time scale is given by
\begin{equation}
    t_{\rm conf} = {\rm max} [t_{\rm diff}, t_{\rm lc}],
\end{equation}
where
\begin{equation}
    t_{\rm diff} \approx \frac{l_b^2}{6D},
\end{equation}
is the diffusion time scale for a spherical blob geometry and $D$ is the diffusion coefficient, see Eq.~\ref{eq:diffusion_coefficient}.
The escape time scale can be estimated as
\begin{equation}
    t_{\rm esc}^p = {\rm min} [t_{\rm conf}, t_{\rm adv}],
\end{equation}
where $t_{\rm adv} = l_b / V$ is the advection escape time scale.

\subsection{Nuclei}
The kinetic equation for nuclei is
\begin{align}\label{eq:transport-nuclei}
    \frac{\partial n_{\varepsilon_A}^{A}}{\partial t} &= -\frac{n_{\varepsilon_A}^{A}}{t_{\rm esc}^A} -\frac{n_{\varepsilon_A}^{A}}{t_{\rm phmes}} 
    -\frac{n_{\varepsilon_A}^{A}}{t_{\rm phdis}}
    + \frac{\partial}{\partial \varepsilon_A} \left[(P_{\rm ad} + P_{\rm BH}^A  + P_{\rm syn}^A) n_{\varepsilon_A}^{A} \right] \nonumber \\ &+ \frac{\partial}{\partial t} \left(n_{\varepsilon_p}^{\rm phmes} + n_{\varepsilon_A}^{\rm phdis}\right) + \dot{n}_{\varepsilon_A}^{\rm inj},
\end{align}
where $t_{\rm esc}^A$ is the nuclear escape time scale, $t_{\rm phmes}$ is the photomeson production interaction time scale, $t_{\rm phdis}$ is the photodisintegration production interaction time scale, $P_{\rm BH}$ is the energy loss rate due to the Bethe-Heitler pair production process, $P_{\rm syn}^A$ is the nuclei synchrotron energy loss rate, $\partial n_{\varepsilon_A}^{\rm phmes} / \partial t$ is the nuclear generation rate from the photomeson production process, $\partial n_{\varepsilon_A}^{\rm phdis} / \partial t$ is the nuclear generation rate from nuclear photodisintegration process, $\dot{n}_{\varepsilon_A}^{\rm inj}$ is the injection rate of primary nuclei.

For the Bethe-Heitler pair production process, we adopt the continuous energy loss approximation similar to the proton case.
The coefficient $\mathcal{A}$ for high-energy nuclei is
\begin{equation}
    \mathcal{A}_{A} = \frac{1}{t_{\rm esc}^A} + \frac{1}{t_{\rm phmes}} + \frac{1}{t_{\rm phdis}} + \frac{\partial (P_{\rm ad} + P_{\rm BH}^A  + P_{\rm syn}^A)}{\partial \varepsilon_p}.
\end{equation}
where the discretization form is 
\begin{equation}
    \mathcal{A}_{A, i} = \frac{1}{t_{\rm esc}^A}\bigg|_i + \frac{1}{t_{\rm phmes}}\bigg|_i + \frac{1}{t_{\rm phdis}}\bigg|_i + \frac{1}{\varepsilon_{i+1/2} - \varepsilon_{i-1/2}}(P_{\rm ad} + P_{\rm BH}^A  + P_{\rm syn}^A)|_{\varepsilon = \varepsilon_i}.
\end{equation}
The coefficient $\mathcal{B}$ is
\begin{equation}
    \mathcal{B}_{A, j \to A, i} = \delta_{j}^{i+1} \frac{1}{\varepsilon_{j+1/2} - \varepsilon_{j-1/2}} (P_{\rm ad} + P_{\rm BH}^A + P_{\rm syn}^A)|_{\varepsilon = \varepsilon_j}.
\end{equation}
The coefficient $\mathcal{C}$ is written as
\begin{equation}
    \mathcal{C}_A = \frac{\partial}{\partial t} \left(n_{\varepsilon_A}^{\rm phmes} + n_{\varepsilon_A}^{\rm phdis}\right).
\end{equation}
Similar to protons, the escape time scale of nuclei depends on both diffusion and advection.

% Don't change these lines
\bsp	% typesetting comment
\label{lastpage}

\end{CJK*}
\end{document}